\documentclass[preprint2]{aastex}

\shorttitle{Variability in PPNs: IV. Four O-rich}

\shortauthors{Hrivnak et al.}

\begin{document}


\title{VARIABILITY IN PROTO-PLANETARY NEBULAE: IV. LIGHT CURVE ANALYSES OF FOUR OXYGEN-RICH, F SPECTRAL-TYPE OBJECTS}

\author{Bruce J. Hrivnak, Wenxian Lu, and Kristie A. Nault\altaffilmark{1}}
\affil{Department of Physics and Astronomy, Valparaiso University,
Valparaiso, IN 46383, USA; bruce.hrivnak@valpo.edu. wen.lu@valpo.edu}

\altaffiltext{1}{Present address: Adler Planetarium, 1300 S. Lake Shore Drive, Chicago, IL 60605, USA; knault@adlerplanetarium.org}

\begin{abstract}

We present new light curves covering 14 to 19 years of observations of four bright proto-planetary nebulae (PPNs), all O-rich and of F spectral type.  
They each display cyclical light curves with significant variations in amplitude.
All four were previously known to vary in light.  
Our data were combined with published data and searched for periodicity.  The results are as follows: IRAS 19475+3119 (HD 331319; 41.0 days), 17436+5003 (HD 161796; 45.2 days), 19386+0155 (101.8 days), and 18095+2704 (113.3 days).  The two longer periods are in agreement with previous studies while the two shorter periods each reveal for the first time reveal a dominant period over these long observing intervals.
Multiple periods were also found for each object.  The secondary periods were all close to the dominant periods, with P$_2$/P$_1$ ranging from 0.86 to 1.06.  
The variations in color reveal maximum variations in T$_{eff}$ of 400 to 770 K.
These variations are due to pulsations in these post-AGB objects.  
Maximum seasonal light variations are all less than 0.23 mag ({\it V}), consistent for their temperatures and periods  with the results of \citet{hri10} for 12 C-rich PPNs.  
For all of these PPNs, there is an inverse relationship between period and temperature; however, there is a suggestion that the period$-$temperature relationship may be somewhat steeper for the O-rich than for the C-rich PPNs. 

\end{abstract}


\section{INTRODUCTION}
\label{intro}

Proto-planetary nebulae (PPNs) are objects in the short-lived evolutionary phase between asymptotic giant branch (AGB) stars and planetary nebulae (PNs).  This transitional phase in the evolution of low- and intermediate-mass 
(0.8$-$8 M$_{\sun}$) 
stars is expected to be several thousand years long, depending upon the mass of the star \citep{blo95}.  This relatively short time scale contributed to the initial difficulty in identifying stars in this stage prior to the availability of data from the {\it InfraRed Astronomical Satellite} ({\it IRAS}), which became available in 1984.   {\it IRAS} allowed the detection of candidate objects based on their mid-infrared emission from circumstellar dust formed during the late-AGB phase.  A good summary of their discovery and properties is given by \citet{kwok00}, while their place in the broader context of post-AGB stars is described by \citet{vanwin03}. 

It soon became apparent that some of the PPNs were oxygen-rich (O-rich) and some were carbon-rich (C-rich).  This was initially based on their mid-infrared spectra as obtained by {\it IRAS} and later by the {\it Infrared Space Observatory} ({\it ISO}).  Those that were O-rich showed amorphous silicate features at 9.7 and 18 $\mu$m, both in emission and absorption \citep{vol91}, and {\it ISO} observations also revealed crystalline silicates at a number of wavelengths between 20 and 43 $\mu$m \citep{mol02}.  Those that were C-rich instead showed infrared aromatic bands between 3 and 12 $\mu$m \citep{hri00b} that are usually identified as due to polycyclic aromatic hydrocarbon (PAH) molecules.  High-resolution visible spectra supported these mid-infrared chemical classifications, with the O-rich showing C to O ratios (C/O) of $\sim$0.5 (summarized in this paper) and the C-rich showing C/O ratios of 1$-$3 \citep{vanwin00,red02}.  The signature of the O-rich or C-rich nature  can also be seen in the gaseous circumstellar nebulae, which can show OH (O-rich) or HCN (C-rich) emission in the centimeter and millimeter radio regions, respectively \citep{lik89,omont93}.

Photometric observations of the brighter PPNs showed that they varied in brightness \citep{ark00,hri00,hri02}.  
Subsequent long-term photometric monitoring has revealed periods for many of them, ranging from 35 to 160 days 
\citep{ark00,ark10,ark11,hri10}. 
When combined with stellar pulsation models \citep[eg.,][]{fok01}, pulsation periods and amplitudes have the potential of yielding information on the mass, luminosity, and stellar structure of these transitional objects.

In this paper, we analyze the light curve variations of four bright O-rich PPNs.  Our observations range from 1994 to 2012.  We also include in our analysis the published complementary photometric observations of these four to increase the data sample.  
This combined sample extends the temporal coverage, fills in gaps in the data, and increases the density coverage.  In these ways it provides an excellent sample from which to search for multiple periods and to compare with the results of our previously published C-rich sample \citep[Paper I]{hri10}.
There is observational evidence that the transition from O-rich to C-rich chemistry occurs for stars with initial masses of approximately 1.5 to 3 M$_{\sun}$ \citep{lat04}.  Thus one might expect different pulsational properties between O-rich and C-rich PPNs, since they result from stars of different masses and luminosities.

\section{PROGRAM OBJECTS}
\label{targets}

The four program objects are listed in Table~\ref{object_list}, along with some basic information.  
The spectra of all four have been classified as F supergiants.  
All have high-resolution spectral abundance studies which not only show that they are O-rich, 
but that they are iron-poor, with [Fe/H] ranging from $-$0.25 to $-$1.1.
Three of them display double-peaked spectral energy distributions \citep[SED;][]{hri88,hri89}.  This is characteristic of PPNs, with the first peak due to the light of the central star, dimmed and reddened by circumstellar dust, and the second peak due to the thermal emission from this dust.
IRAS 19386+0155 displays only a single peak in its SED.
Below is a brief background description of each.

\placetable{object_list} 

\subsection{IRAS 17436+5003}
Initial attention to the unusual properties of IRAS 17436+5003 (HD 161796, V814 Her) was drawn by \citet{bid81}, who listed it among a small group of four supergiants located at high galactic latitude. 
 Following the launch of {\it IRAS}, \citet{partha86} noted its large infrared excess and suggested that it might be a post-AGB object, as did \citet{hri89}, who included it in a multi-wavelength study of a eight PPN candidates of F$-$G spectral types. 
Imaging with the {\it Hubble Space Telescope} ({\it HST}) revealed a small elliptical nebula with size of 
4.4$\arcsec$ $\times$ 2.5$\arcsec$ surrounding the bright star \citep{ueta00}.
The mid-infared {\it ISO} spectrum shows the broad amorphous silicate emission features at 9.7 and 18 $\mu$m and additional crystalline silicate features in the 15$-$70 $\mu$m region \citep{mol02}.

\subsection{IRAS 18095+2704}
The basic properties of IRAS 18095+2704 (V887 Her) were initially discussed by \citet{hri88}, who identified the {\it IRAS} source with a then un-cataloged 10th magnitude star.
Its {\it HST} image shows an elliptical or perhaps bipolar nebula with a size of 1.9$\arcsec$ $\times$ 1.2$\arcsec$ surrounding the bright star \citep{ueta00}.
The mid-infared {\it ISO} spectrum displays silicate emission at 9.7 and 18 $\mu$m, and weak emission features in the region from 10$-$27 $\mu$m that are attributed to crystalline silicates \citep{hri15b}.  
This object is an OH maser \citep[OH 53.8+20.1]{eder88}.

\subsection{IRAS 19386+0155}
The basic properties of IRAS 19386+0155 (V1648 Aql) were initially discussed by \citet{veen89} in their study of post-AGB transitional objects. 
\citet{per04} carried out a visible and infrared spectroscopic study to determine the chemical composition of the star and to model its envelope.
The SED shows only one broad peak reaching maximum energy in the region from 4.5 to 25 $\mu$m.  This seems to indicate a relatively obscured and reddened central star.
It has not been observed with {\it HST}, but it has been observed and resolved in the mid-infrared, with a size of 2.6$\arcsec$ and possessing a bright core and extended halo \citep{lag11}. 
The mid-infrared {\it ISO} spectrum shows the 18 $\mu$m silicate emission feature and features in the 10 $\mu$m region that we interpret as self-absorption in the middle of a silicate emission feature. \citep{hri15b}.
OH maser emission has been detected \citep{lew00}.

\subsection{IRAS 19475+3119}
The basic properties of IRAS 19475+3119 (HD 331319, V2513 Cyg) were broadly discussed by \citet{garlar97}.
{\it HST} imaging shows that the nebula has an interesting, quadrupolar shape surrounded by a faint halo \citep{sah07}. 
The mid-infrared {\it ISO} spectrum shows some features identified as crystalline silicates \citep{sar06}.

\section{PHOTOMETRIC OBSERVATIONS}

Photometric observations were carried out at the Valparaiso University Observatory (VUO) using the 0.4-m telescope and a CCD detector.  They began in 1994 and continue through the present time; those reported in this study run through 2012, except for IRAS 19386$+$0155, for which we stopped observations in 2007.  
Initial observations were made primarily through the {\it V} filter, with some occasionally through the {\it R$_C$} filter, but the frequency of {\it R$_C$} observations increased and since 2000 they they have been made together.
Unfortunately a problem developed with the {\it V} filter which caused us to reject the {\it V} data from the 2000 and 2001 seasons and part of 2002.
This led to a gap in each of the {\it V} light curves.  
Beginning in 2007, {\it B} observations were also made for all of the sources except IRAS 19386+0155. 
In 2008 we upgraded from a Photometric Star I CCD, with 
a field of view of 8$\farcm$6 $\times$ 5$\farcm$7, to an 
SBIG 6303 CCD, with 
a field of view of 17$\farcm$9 $\times$ 11$\farcm$9.
Filter sets were used with each 
to match the standard Johnson {\it B} and {\it V} and Cousins {\it R$_C$} systems. 
The observations were reduced with IRAF\footnote{IRAF is distributed by the National Optical Astronomical Observatory, operated by the Association for Universities for Research in Astronomy, Inc., under contract with the National Science Foundation.}, using standard procedures to remove cosmic rays effects, subtract the bias, and flat field the images. 
Aperture photometry was carried out with an aperture of 11$\arcsec$ diameter.

Given the frequently partially-cloudy nights at our site, we carried out a program of differential photometry.  Three comparison stars were used for each of the program stars.  Ideal comparison stars would be a good match in color and brightness, but that was not achievable in these cases due to the brightness of the program stars and the size of the field of view of our CCD detectors.  
The comparison stars used are listed in Table~\ref{phot_std}.  
When we changed to the larger format CCD, we were able to include new, brighter comparison stars C$_2$ and C$_3$ for IRAS 17436+5003, and these are the ones that we have listed in Table~\ref{phot_std}.
For each PPN, the constancy of the primary comparison star, C$_1$, was examined with respect to the other two comparison stars.
The precision of the observations is not as high as we would like due to the large differences in brightness between the program and comparison stars.
This restricted the integration times to avoid non-linearity in the program stars, but consequently lead to less precision in the comparison star observations. 
We find the comparison stars to be reasonably constant over time, to within $\pm$0.015 mag, much smaller than the variations found in the program stars. 
Comments about individual comparison stars are found in the footnotes to Table~\ref{phot_std}. 
In Table~\ref{tab_stat} are listed the numbers of observations of the program stars in each filter, along with the statistical uncertainties in the differential magnitudes.  
The typical precision in the differential magnitudes is $\pm$0.006 mag. 

\placetable{phot_std}

\placetable{tab_stat}

Standardized photometry of these program star fields was carried out on two different nights at the VUO, 
and the standard magnitudes are listed in Table~\ref{phot_std}.
The differential magnitudes of the PPNs were also transformed from the instrumental to the standard magnitude systems.  
Standardized {\it V} and {\it R$_C$} differential photometry for the four program stars from 1994 to 2007 is listed in Table~\ref{diffmagVR_std}, except for the 2007 observing sets that included {\it B} also.
Those are listed in Table~\ref{diffmagBVR_std} along with {\it BVR$_C$} observations made with the newer CCD beginning in 2008.  

\placetable{diffmagVR_std}

\placetable{diffmagBVR_std}

\section{VUO LIGHT CURVES}
\label{vuo_lc}

\subsection{IRAS 17436+5003}

\begin{figure}\figurenum{1}\epsscale{1.0}
\plotone{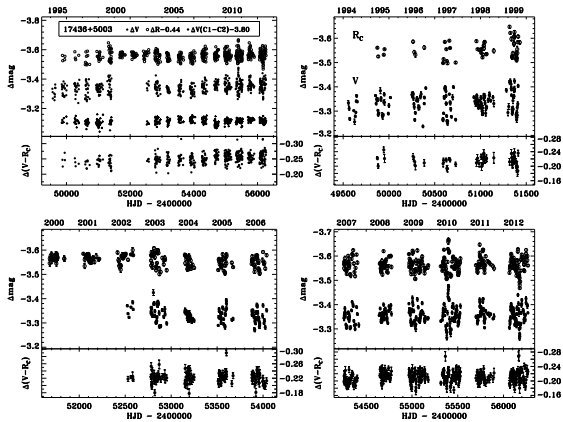}
\caption{Plots showing the differential light and color curves of IRAS 17436+5003 obtained at the VUO from 1993$-$2012.  In the upper left panel are the entire curves and in the three other panels the curves are shown on expanded scales covering six or seven years each.  The error bars are included with the data in these expanded plots.  
In the upper left panel is also shown the differential V light curve of the primary (C$_1$) with respect to the secondary (C$_2$) comparison stars on the same scale as the PPN light curves. 
The absence of {\it V} data from 2000 to part of 2002 is due to a filter problem.
Zero-point offsets are added to conveniently display all three light curves on the same plot. 
\label{17436_oldvr}}\epsscale{1.0}
\end{figure}

\begin{figure}\figurenum{2}\epsscale{0.7}
\plotone{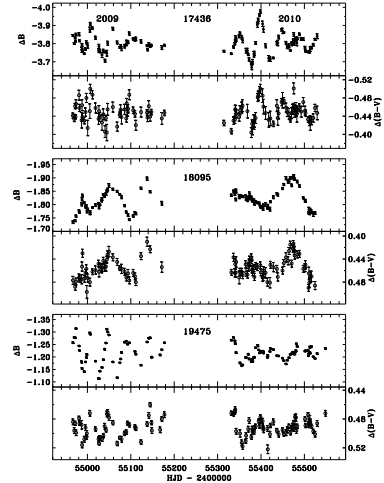}
\caption{Plots showing samples of the differential {\it B} and ({\it B$-$V}) curves from the 2009 and 2010 seasons obtained at the VUO. The changing cyclical behavior of the light and color curves can be seen, as can the correlation between brightness and color.  Error bars are included. 
\label{LC_newB}}\epsscale{1.0}
\end{figure}

The {\it VR$_C$} light and color curves of IRAS 17436+5003 are displayed in Figure~\ref{17436_oldvr}.
The light curves possess a total peak-to-peak variation of 0.23 mag in {\it V} and 0.18 mag in {\it R$_C$}, with a relatively large variation in the yearly amplitudes and a slight variation in the yearly means. 
This variability is superimposed perhaps on a slight, non-monotonic trend of increasing brightness of $\sim$0.04 mag ({\it V}) over the 19 years of our observations.  
The peak-to-peak range within a season varies greatly, from 0.05 (2004) to 0.21 (2010) mag in {\it V} and from 0.06 (2004) to 0.18 (2010) mag in {\it R$_C$}, with an average ratio of the seasonal range of {\it R$_C$} to {\it V} of 0.90.
The variations are even larger in {\it B} and ({\it B$-$V}),
with an average ratio of the seasonal {\it B} to {\it V} of 1.36 over the six years in which we have {\it B} observations. 
Samples of the {\it B} light and the ({\it B$-$V}) color curves are shown in Figure~\ref{LC_newB}.
The differential comparison star measurements $\Delta${\it V}(C$_1$$-$C$_2$) show a relatively large amount of scatter (see Fig.~\ref{17436_oldvr}) due to the faintness of the comparison stars relative to IRAS 17436$+$5003 ({\it V}=7.0) and, from 1994$-$2007, the faintness of the initial C$_2$ star relative to C$_1$.  There is evidence of a general increase in brightness of C$_1$ by 0.02 mag ({\it V}) from 1994 to 2007; this is relatively small and would not affect our main results.
Inspection of the seasonal light curves of IRAS 17436+5003 shows in general a cyclical variation with a period in the range of $\sim$40$-$50 day.  
At times, however, it appears as though the variation has gone away, such as at the ends of the 2009 and 2011 observing seasons (e.g., Fig.~\ref{LC_newB}).

\placefigure{17436_oldvr}

\placefigure{LC_newB}

\begin{figure}\figurenum{3}\epsscale{1.00}
\plotone{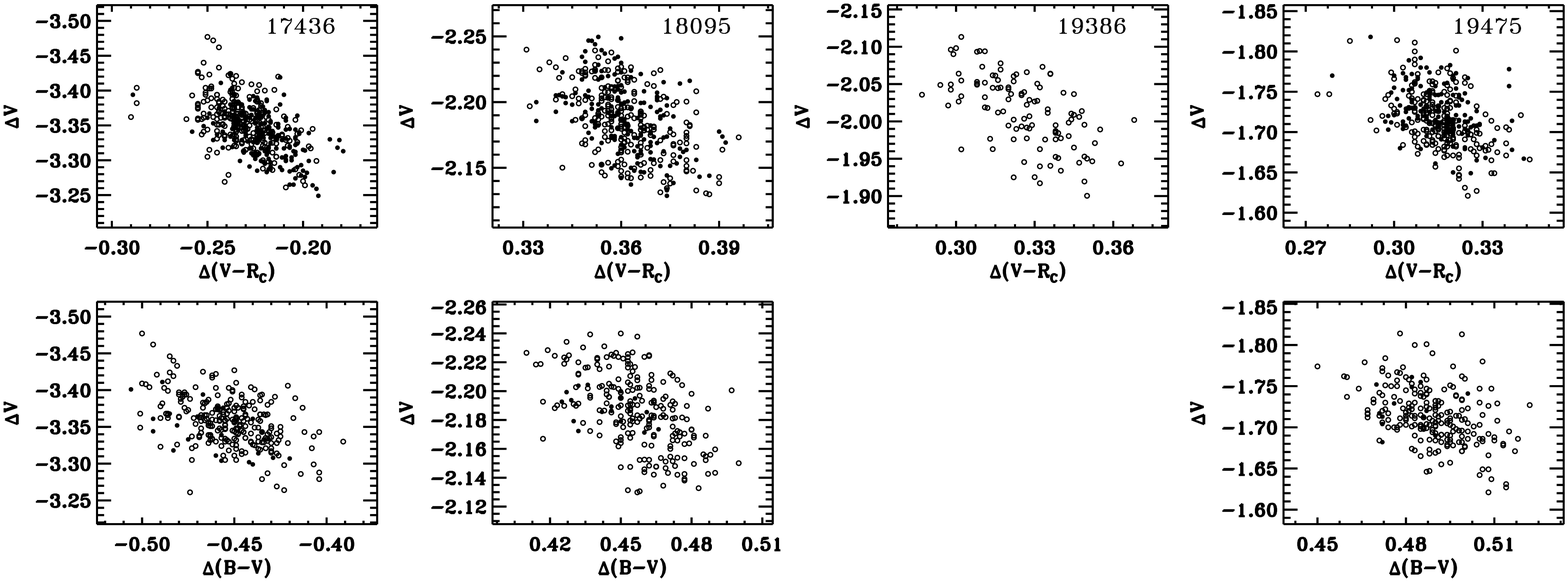}
\caption{Plots showing the change in color with change in brightness.
A clear trend is seen, with the objects being redder when fainter.  The observations with the initial VUO CCD are plotted as filled circles and with the newer CCD as open circles.  
For IRAS 18095+2704, an adjustment in ({\it V$-$R$_C$}) of 0.01 mag has been applied to correct for the small offset in color between the old and new CCD systems, as described in the text.
\label{LC_color}}\epsscale{1.0}
\end{figure}

The ({\it V$-$R$_C$}) color of the system shows a variation of 0.07 mag peak-to-peak (except for four outlying points), with most of the data within a range of 0.05 mag; the ({\it B$-$V}) color range is larger, reaching 0.12 in 2010, with most of the data within a range of 0.07  mag.  An inspection of the light and color curves shows that, in general, there is a trend of color with brightness, being redder when fainter. 
This can perhaps be seen more clearly in a plots of $\Delta${\it V} versus $\Delta$({\it V$-$R$_C$}) and $\Delta${\it V} versus $\Delta$({\it B$-$V}), as shown in Figure~\ref{LC_color}.

\placefigure{LC_color}

\subsection{IRAS 18095+2704}

\begin{figure}\figurenum{4}\epsscale{1.0}
\plotone{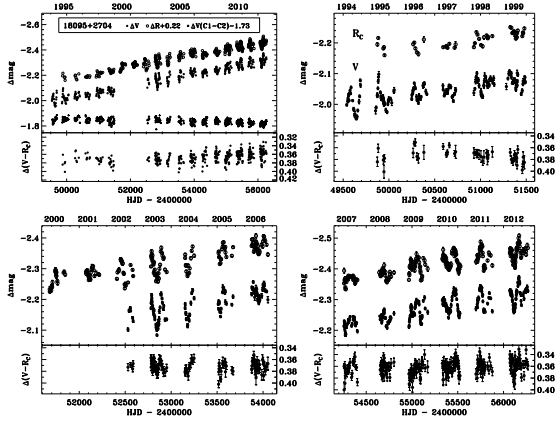}
\caption{Plots showing the differential light and color curves of IRAS 18095+2704 obtained at the VUO, plotted similar to Figure~\ref{17436_oldvr}.  
\label{18095_oldvr}}\epsscale{1.0}
\end{figure}

The {\it VR$_C$} light curves for IRAS 18095+2704 are shown in Figure~\ref{18095_oldvr}.
They display a general trend of increasing brightness over the two decades of observations, increasing by 0.30 mag in {\it V} from 1994$-$2012 and 0.26 mag in {\it R$_C$} from 1995$-$2012.
At the same time, there is very little change in the seasonal ({\it V$-$R$_C$})
color of the system, $<$$-$0.01 mag, apart from what appears to be a slight zero-point shift of $-$0.01 mag that arose when we changed from the initial to the newer CCD in 2008.  
We assume that this difference arises from some unresolved imprecision at this level in the standardization of the two systems for this star. 

\placefigure{18095_oldvr}

Superimposed on this brightening trend is a clear cyclical variation of greatly varying amplitude.
The amplitude (peak-to-peak) varies in {\it V} from 0.05 (1997) to 0.14 mag (2003) and 
in {\it R$_C$} from  0.02 (1997) and 0.04 (1999) to 0.11 mag (2003).   
For years in which we have good coverage with both filters, the average ratio of the amplitude in the two ({\it R$_C$} compared with {\it V}) is 0.86.  In the six years in which we observed consistently with all three filters, the average ratio of the amplitude in {\it B} to {\it V} is 1.29.  
Samples of the more recent {\it B} and color data, showing the cyclical variations, are shown in Figure~\ref{LC_newB}. 

Looking at the ({\it V$-$R$_C$}) color (Fig.~\ref{18095_oldvr}), it can be seen that in general it follows the cyclical variation, being redder when fainter.  
This is perhaps seen even more clearly in the ({\it B$-$V}) data (Fig.~\ref{LC_newB}).
Plotted in Figure~\ref{LC_color} are the $\Delta${\it V} versus  $\Delta$({\it V$-$R$_C$}) and $\Delta${\it V} versus  $\Delta$({\it B$-$V}) data sets, with the monotonic trend in $\Delta${\it V} brightness first removed using a low-order polynomial fit.  
The ({\it V$-$R$_C$}) color between the two different CCD systems have been adjusted by the small offset of $-$0.01 mag mentioned above.
There is a clear linear trend, with the object being redder when fainter.
The range in $\Delta$({\it V$-$R$_C$}) is 0.06 mag, with most within a range of 0.04 mag,
the range in $\Delta$({\it B$-$V}) is 0.09 mag, with most within a range of 0.05 mag.

\subsection{IRAS 19386+0155}

\begin{figure}\figurenum{5}\epsscale{1.0}
\plotone{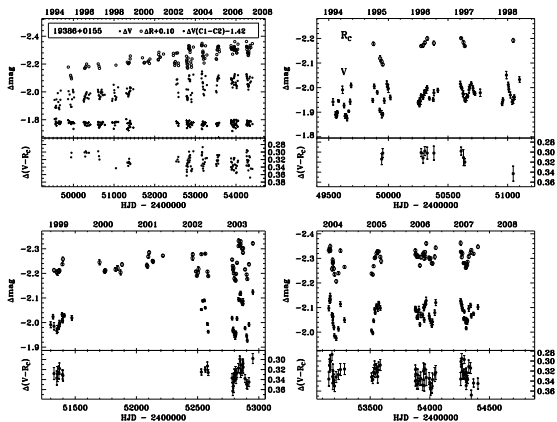}
\caption{Plot showing the differential light and color curves of IRAS 19386+0155 obtained at the VUO from 1993$-$2007, plotted similar to Figure~\ref{17436_oldvr}.  
\label{19386_oldvr}}\epsscale{1.0}
\end{figure}

IRAS 19386+0155 also displays a general increase in brightness over the interval of observation (1994$-$2007), as seen in Figure~\ref{19386_oldvr}. 
It shows an increase of 0.13 mag in {\it V} over 14 years.
During this time, there is a suggestion that the {(\it V$-$R$_C$}) color was bluer by 0.02 mag in the first several years, but this result is uncertain since it is based on only a small number of {\it R$_C$} observations in the years 1995-1997 compared with the later observations from 1999 and from 2002 through 2007.

\placefigure{19386_oldvr}

A cyclical variation is also seen in the seasonal data, with a varying amplitude.  
The amplitude ranges in {\it V} from 0.07 (1996, 1997, 1999) to 0.20 mag (2003). 
There are many fewer {\it R$_C$} observations in the early years, but from 2003 through 2007, 
the ratio of amplitudes ({\it R$_C$} compared with {\it V}) is 0.84. 

When comparing the seasonal ({\it V$-$R$_C$}) color variations with the seasonal cyclical variations in brightness (Fig.~\ref{19386_oldvr}), one can see a general pattern of the object being redder when fainter.  
This can be seen clearly in Figure~\ref{LC_color}, 
following the removal of the brightness trend in {\it V} using a low-order polynomial fit.  
The range in color is small, $\sim$0.06 mag (ignoring three outliers).

\subsection{IRAS 19475+3119}

\begin{figure}\figurenum{6}\epsscale{1.0}
\plotone{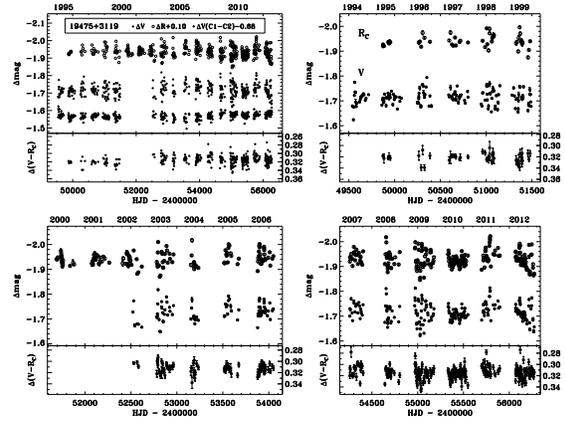}
\caption{Plot showing the differential light and color curves of IRAS 19475+3119 obtained at the VUO, plotted similar to Figure~\ref{17436_oldvr}.  
\label{19475_oldvr}}\epsscale{1.0}
\end{figure}

The light curves of IRAS 19475+3119 show no long-term trends in brightness, although they do show some small variations in the mean seasonal brightness and some changes in the seasonal range in brightness.  These are seen in Figure~\ref{19475_oldvr}.
The seasonal range in {\it V} brightness varies from lows of 0.07 (1995, 1997) and 0.08 (2010) mag to 
highs of 0.19 (2009) and 0.17 (2003) mag.  
The range in {\it R$_C$} is typically less, with an average ratio ({\it R$_C$} compared with {\it V}) of 0.94, and the range in {\it B} is typically more in the six years in which it is observed, with an average ratio ({\it B} to {\it V}) of 1.29.
Visual inspection of the seasonal light curves shows cyclical variations in brightness, and in several of the seasons (1996, 1998, 2003, 2008, 2009, and 2010) one can discern a cycle length of 30$-$50 day (see Fig.~\ref{LC_newB}). 

\placefigure{19475_oldvr}

The overall color of the system does not vary much, with the $\Delta$({\it V$-$R$_C$}) value mostly within the range of 0.035 mag and $\Delta$({\it B$-$V}) value mostly within the range of 0.04 mag.
When one examines the seasonal light and color curves, one can see a trend, with the object fainter when redder 
(see Fig.~\ref{LC_newB}). 
Such a trend can be seen clearly when looking at the overall brightness-color curves, 
as shown in Figure~\ref{LC_color}.

\section{COMBINED LIGHT CURVES \& PERIOD ANALYSES}
\label{PeriodStudy}

All four of these PPNs are known to be variable in light and have been included in prior studies by other investigators.  This is particularly true of the brightest of these sources, IRAS 17436+5003, which has been observed by several investigators, particularly Fernie.  
The other three sources have been observed extensively by Arkhipova and collaborators, as described below.  Their {\it UBV} observations are carried out with a single-channel photoelectric photometer using a 27$\arcsec$ aperture, and they tied their observations to the {\it UBV} system by observations of standard stars.
Data are also available for two of the objects in the {\it V} band from the All Sky Automated Survey 
\citep[ASAS;][]{poj02}\footnote{http://www.astrouw.edu.pl/asas/}.    

The availability of these additional observations affords an opportunity to enlarge the data sample, either in time or density or both, with the promise of better documented light curves.  These in turn can yield better determined results for the period analysis, including the determination of multiple periods.
This is, of course, predicated on the ability to accurately combine the different data sets.
All of the investigators made differential observations, as we did, and to combine them requires accurate standardized measurements of the various comparison stars.  Since different investigators use different comparison stars and different aperture sizes, this can potentially lead to slight zero-point offsets in the magnitudes in the different data sets.
To correct for such offsets, we compared the various data sets where they overlapped in time, to determine and correct for any such offsets.  These are discussed for each star.
These individually-derived offsets are well established, based on a minimum of 18 nights, and are listed in Table~\ref{LC_offset}, along with access to the published data sets.
We briefly discuss these other studies, star by star, along with the resulting combined light curves, and 
we then proceed to the period analyses of each.

\placetable{LC_offset}

Period studies were carried out for each of the four PPNs using the period-finding program 
Period04 \citep{lenz05}.  This uses a Fourier transform, and it easily allows for the investigation of multiple periods.  We began with the large, combined {\it V} data sets and then examined the data in subsets.
For IRAS 18095+2704 and 19386+0155, the general trends were first removed.
The light curves each show some variations in the yearly means.  We first searched the data sets for periodicity with these included, in case there were longer-term periodicities that these represented.  
For none of these PPNs was a significant longer-term periodicity found, so we then normalized, or adjusted, the yearly data by the yearly means and then did the final period analyses on the adjusted data sets.

\subsection{IRAS 17436+5003}
\subsubsection{Combined light curves}

Due to its brightness and its unusual distinction of being a high-latitude object with a supergiant spectrum, IRAS 17436+5003 (HD 161796)  became the target of many variability studies even before it was identified as a PPN.  
The most extensive of these was by Fernie and collaborator \citep{fern83,fern86,fern89,fern90a, fern90b,fern91,fern93,fern95}, who carried on a long series of observations including most years from 1980 through 1998.  
They always include {\it V} observations and from 1986 to 1998 the observations are {\it UBV}.
Beginning in 1986, the observations were made with an automated photoelectric telescope at a good site, and the number of annual observations increased significantly.

We combined with our {\it V} observations the long and consistent data set of Fernie's.   
In comparing the data of Fernie's with ours during the overlap from 1984 to 1998, we find that there are 22 dates on which we observed on the same night, and from these we find an average offest of 0.000 $\pm$0.002.  Thus there is no offset to apply.  
We extended the coverage slightly by including the data by \citet{percy81} from 1979$-$1980;
comparing these with those of Fernie from 1980 indicated no offset. 

\begin{figure}\figurenum{7}\epsscale{0.72}
\plotone{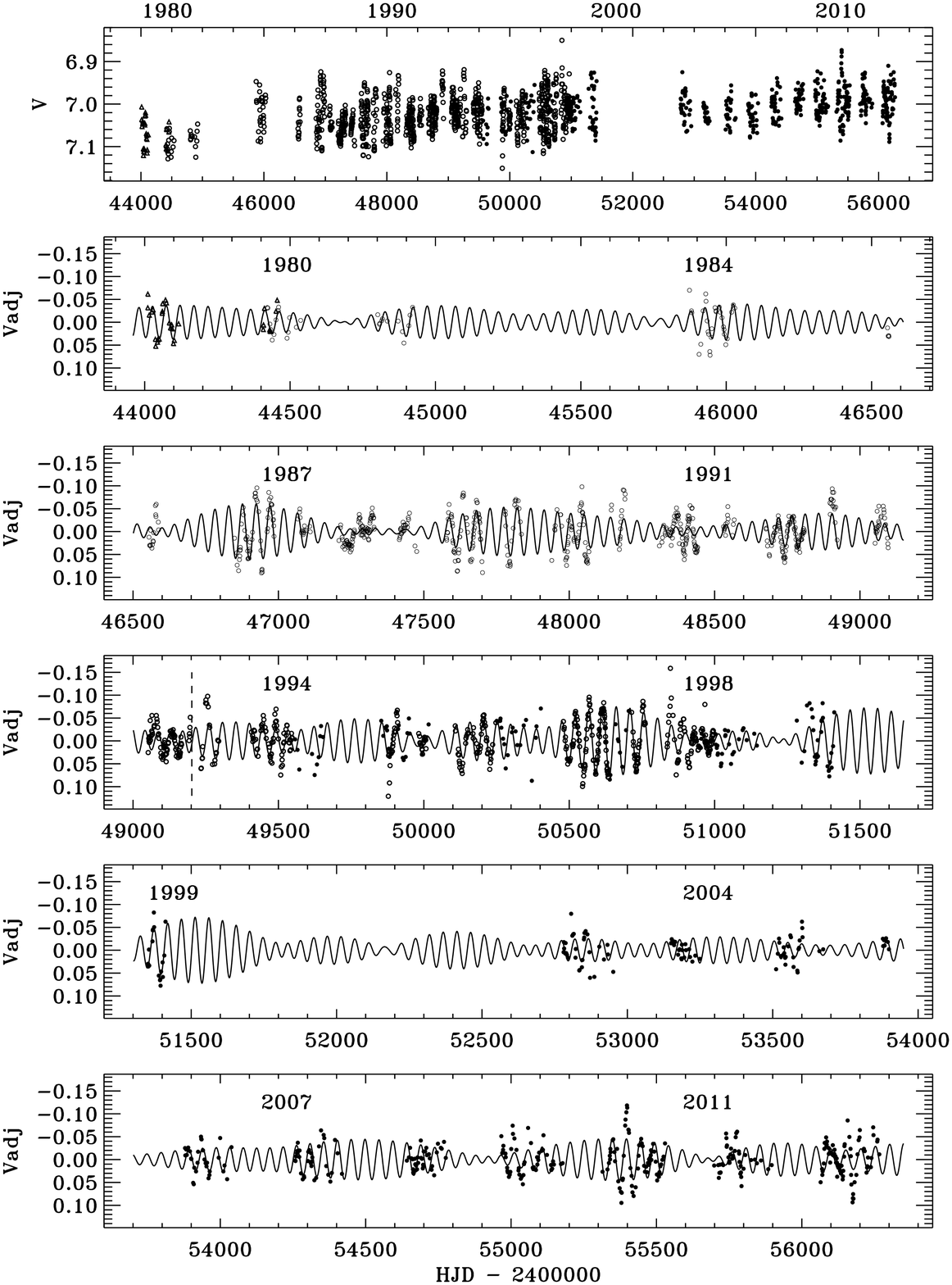}
\caption{Top panel: Combined {\it V} light curve of IRAS 17436+5003, with our data shown as filled circles, the data of Fernie shown as open circles, and the data of Percy \& Welch shown as filled triangles.
Bottom five panels: The seasonally-adjusted {\it V} light curve fitted with the six periods and amplitudes listed in Table~\ref{lc_per}.  The fit is generally good throughout except for the most recent data (bottom panel).  Note that there is some repetition of the data and the fitted curves at the edges of the five lower panels.
The vertical dashed line shows the time when published high-resolution spectra were taken.
\label{17436_fit}}\epsscale{1.0}
\end{figure}

The combined {\it V} light curve is displayed in Figure~\ref{17436_fit}. 
The agreement of the different data sets is good in the regions of overlap and very usefully increases the temporal coverage. 
The light curve shows a general increase in brightness of 0.04 mag from 1984 to 2012, but the increase is not strictly monotonic.  The yearly amplitude is seen to vary throughout this interval by a factor of three to four.  

\placefigure{17436_fit}

\subsubsection{Period analyses}

The combination of our data with those of Fernie and Percy \& Welch gives us a relatively-long time interval of 34 years of data for IRAS 17436+5003. 
This affords a good opportunity to not only find the best period(s) for the entire data set, but to also examine the observations in smaller time intervals to investigate period changes.  The data sets were first adjusted to the mean brightness of each year.
Years with fewer than ten data points were not included.

\begin{figure}\figurenum{8}\epsscale{1.10} 
\plotone{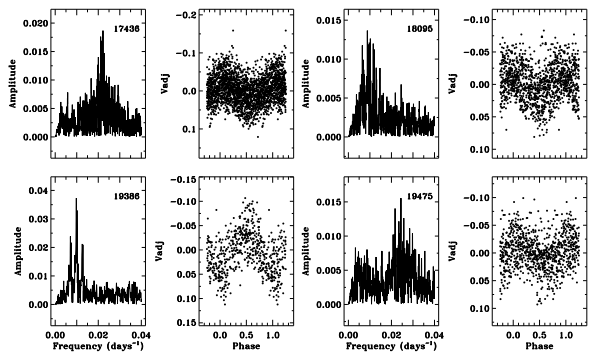}
\caption{The frequency spectrum of the combined and seasonally-adjusted {\it V} light curves of the four targets,
together with their associated phase plots based on the frequency peaks (P$_1$).
The presence of secondary period peaks in the data is evident.
\label{freqspec}}
\epsscale{1.0}
\end{figure}

The analysis of the entire {\it V} data set (set A) revealed in several closely-spaced frequencies, as can be seen in Figure~\ref{freqspec}.  The analysis revealed dominant periods of P$_1$ = 45.15 $\pm$ 0.01 days, P$_2$ = 47.36 days, and P$_3$ = 46.75 days.  There are in total eight significant periods in the data; we have listed the first six in Table~\ref{lc_per}.  The data, phased to P$_1$ = 45.15 days, are also displayed in Figure~\ref{freqspec}.  They show a reasonably good phased light curve, considering the observed range in amplitude seen in the light curve.  
The fit of the periods and amplitudes to the normalized light curve is remarkably good for most of the time interval, as can be seen in the lower five panels of Figure~\ref{17436_fit}.  It is only in the last decade, 2003$-$2012, that the fit in some seasons is not good.  Some of this may be due to the observed complexity of the light variations and some may be due to changes in the periods.  
Comments about the light curve complexities and period changes were frequently made by Fernie.  Fernie found dominant periods ranging from 38 to 62 days when he examined the observations in one or two year intervals, although most were in the range of 40$-$46 days.
Fernie also drew attention to some epochs in which the star appeared to not vary \citep{fern93b}.
Some of this complexities seem to be resolved by our inclusion of multiple periods.

\placefigure{freqspec}

\placetable{lc_per}

The data were then examined in subsets and the results are also listed in Table~\ref{lc_per}.  The analysis of the Fernie and Percy \& Welch data (set B, 1979$-$1998) results in  P$_1$ = 45.08 $\pm$ 0.01 days, P$_2$ = 46.54 days, and P$_3$ = 47.45 days; these are nearly the same as the first three periods seen in the entire data set.  This is perhaps not surprising, since their data comprise two-thirds of the number in the entire data set.
The analysis of the complete VUO data (set C, 1994$-$2012) results in  P$_1$ = 47.30 $\pm$ 0.02 days, which agrees with P$_2$ in the entire data set (A), and a value for P$_2$ that agrees with P$_4$ in the entire data set.
Similar results were found in the VUO {\it R$_C$} data (set D, 1997$-$2012).
We also looked at subsets of the VUO {\it V} data in five or six year intervals.
The 1994$-$1999 data (set E) has  P$_1$ = 44.09 and P$_2$ = 42.38 days; the second of these agrees with P$_3$ of the entire VUO {\it V} data set (C).
The 2003$-$2007 data (set F) has  P$_1$ = 41.47 and P$_2$ = 56.87 days; this value of P$_1$ does not appears in any of the other data sets.
The 2008$-$2012 data (set G) has  P$_1$ = 47.42 and P$_2$ = 57.70, and P$_3$ = 38.14 days; P$_1$ agrees with the value found for P$_1$ in the complete VUO data set (C) and P$_2$ is similar to the value of P$_2$ found in the 2003$-$2007 data set.

Thus, while there are several common periods found in many of the data sets, particularly 45.1 and 47.4 days, there are also values of {\it P} that are found in only certain time intervals (sets E, F).  This suggests that there are indeed changes in the dominant periods with time.
However, there does not appear to be a monotonic increase or decrease in the period over time, as seen by the comparison of the P$_1$ values in subsets E, F, and G.
One can also meaningfully compare the ratio of the two dominant periods, particularly in the longer data sets (A, B, C, D).  They are all close in value, ranging from 1.03 to 1.06, with an average value of 1.05.\\

\subsection{IRAS 18095+2704}
\subsubsection{Combined light curves}

IRAS 18095+2704 was observed by \citet{ark93,ark00,ark10} for 19 years, beginning with only a few observations in 1990$-$1992 but then more frequently through 2008.  
They quote an uncertainty of $\pm$0.01 mag in {\it B} and {\it V}.
Their observations show a clear trend of increasing brightness, as we found, with an increase of 
0.37 mag in {\it V}, 0.35 mag in {\it B}, and 0.32 mag in {\it U} over 19 years, 
with the ({B$-$V}) color staying approximately constant.
Examining some older photographic observations, they find that this brightness trend extends back to the earliest measurements from the mid-1930s.
From their 2000$-$2008 observations, they find a period of 109$\pm$2 days with a semi-amplitude of 0.03 mag, and they note that during this cyclical variation the system is bluer when brighter.

To combine our {\it V} data with theirs, we first compared the two data sets.  There are 24 nights in common between the two data sets, with a systematic difference (VUO $-$ Arkhipova) of $-$0.054 mag and another 50 nights in which observations from the two data sets were made on successive nights, with a difference of $-$0.042.  This yielded a weighted value of $-$0.048 mag, with those on successive nights receiving half weight.  
Possible reasons for this offset include uncertainties in standardization of the measurements, including the comparison star, and perhaps, due to their larger aperture, an unseen star in their comparison star aperture.

As part of the ASAS sky survey, 69 observations of IRAS 18095+2704 with a typical uncertainty of $\pm$0.03 mag were made during 2003 and part of 2004. The 2003 data show a well-delineated cyclical pattern over almost two cycles with a period of $\sim$105 days. However, we did not combine these data with ours. 

\begin{figure}\figurenum{9}\epsscale{1.0}
\plotone{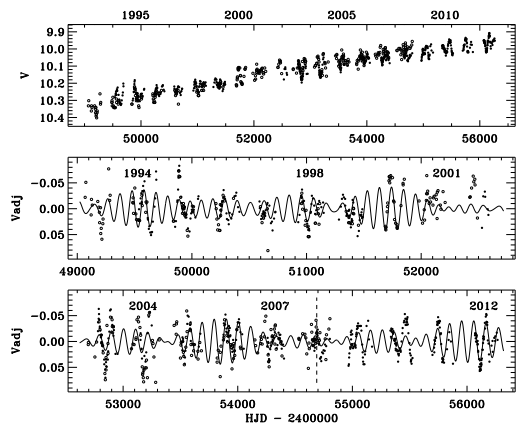}
\caption{Top panel: Combined {\it V} light curve of IRAS 18095+2704, with our data shown as filled circles and the data of Arkhipova et al. shown as open circles.
Bottom two panels: The trend-removed, seasonally-adjusted {\it V} light curve on an expanded scale, fitted with the four periods and amplitudes listed in Table~\ref{lc_per}.  The fit is reasonable on most seasons given the complex nature of the light curve.  In some seasons, the periods looks like they agree but not the fixed amplitudes.  The vertical dashed line shows the time when published high-resolution spectra were taken.
\label{18095_fit}}\epsscale{1.0}
\end{figure}

A combined {\it V} light curve was formed by adding the Arkhipova et al. data set from 1993 to 2008 
to ours, using the offset determined.  This is displayed in the top panel of Figure~\ref{18095_fit}, and shows an increase in brightness of 0.40 mag over this 20-year interval.
To prepare the combined data set for analysis, we first removed the trend in each data set separately using a low-order polynomial, and then adjusted the seasonal light curves to the same mean levels.  This adjusted {\it V} light curve is shown in the bottom panel of Figure~\ref{18095_fit}.  
Close examination of the seasonal light curves shows good agreement between the two data sets and serves to better document the cyclical variability.

\placefigure{18095_fit}

\subsubsection{Period analyses}

We began with a period analysis of the combined, adjusted {\it V} light curve of IRAS 18095+2704, which covers the longest time interval, 1993 to 2012, and has the largest number of data points (set A).  The frequency spectrum showed several relatively strong peaks, and the subsequent analysis indicated multiple periods in the data.  The strongest period was P$_1$ = 113.3 $\pm$ 0.1 days, with additional periods of 97.9, 103.0, and 108.4 days.  The frequency power spectrum and the phase plot with P$_1$ are shown Figure~\ref{freqspec}, and the presence of these additional periods is seen in the frequency spectrum.
Six significant periods are found in the data, and these are listed in Table~\ref{lc_per}. 

In Figure~\ref{18095_fit} is also shown the fit of the first four periods, along with their associated amplitudes and phases, to the observed light curve.  The fit is reasonably good, given the complex nature of the light curve and the long interval of observations (20 years), and improves slightly with the inclusion of the two additional significant periods.  

Subsets of this {\it V} light curve and of some of the light curves with other filters were also investigated.
We began by dividing the combined {\it V} data set into two ten-year time intervals.
For the time interval 1993$-$2002 (set B), the analysis reveals P$_1$ = 98.2 days, close to P$_2$ in the full {\it V} data set, and an additional significant period of 107.5 days (close to P$_4$ in the full data set).
For the interval 2003$-$2012 (set C), there is a strong period at P$_1$ = 111.8 days, close to P$_1$ in the full {\it V} data set, and additional periods of 110.2, 103.1 (close to P$_3$ in the full data set), 68.3, and 97.2 days; the periods of 111.8 and 110.2 in the 2003$-$2012 {\it V} data set are close to and beat against one another.
Note that a formal analysis of the ASAS 2003$-$2004 {\it V} light curve results in a period of 110 $\pm$1 days.
It can be seen that the same periods are not dominant throughout the 20-year observing interval, although almost all of those in the two 10-year intervals are found in the combined data set.
Next we investigated the {\it B} light curve, formed by removing the trends from the Arkhipova  et al. and the VUO data, including the determined {\it B} offset, and adjusting to the individual seasons.
Excluded were two years with fewer than 10 observations.
This data set (D) yielded periods in agreement with those found for the combined {\it V} light curve, P$_1$ = 113.5 days and additional periods of 103.0, 154.5, 97.9, and 116.3 days, and with larger amplitudes than found for the {\it V} analysis. 
We also investigated the VUO {\it R$_C$} light curves from 1998$-$2012 (set E) and 2003$-$2012 (set F).  Some of the same periods appear, but not exactly, and the {\it R$_C$} light curves have smaller amplitudes.
These periods are also recorded in Table~\ref{lc_per}.
Based on the combined, full {\it V} light curve (set A), we find a ratio of P$_2$/P$_1$ = 0.86.

\subsection{IRAS 19386+0155}
\subsubsection{Combined light curves}

\citet{ark93,ark00,ark10} also observed IRAS 19386+0155 for 19 years, again beginning with only a few observations in 1990$-$1992 but then more frequently through 2008.  
They quote an uncertainty of $\pm$0.01 mag in {\it B} and{\it V}.
Their observations show a clear trend of increasing brightness with time.
They determined an increase in {\it V} of 0.2 mag but find that surprisingly the ({B$-$V}) color gets redder by 0.1 mag during this time.
From the 2000$-$2008 observations, they derived a dominant period of 102 day and additional periods of 98 and 106, yielding a period ratio P$_2$/P$_1$ = 0.96. They note that during this cyclical variation the system is bluer when brighter, as we found.

To combine our {\it V} data, we compared the two data sets and found that there are 12 nights in common, with a systematic difference (VUO $-$ Arkhipova) of $+$0.060 mag and another 26 successive nights of observations, with a difference of $+$0.061.  This yielded a weighted value of $+$0.060 mag, with those on successive nights receiving half weight.  

IRAS 19386+0155 was also observed with in the ASAS sky survey from 2002 through 2009.
We examined the data set of 766 {\it V} observations, 
which had a average uncertainty of $\pm$0.035 mag.  
Comparing their data with ours on similar nights lead to an offset of $-$0.055 mag.
Given the significantly larger uncertainty in the data, we chose not to combine them with our data; we did, however, investigate them separately for periodicity.

\begin{figure}\figurenum{10}\epsscale{1.0}
\plotone{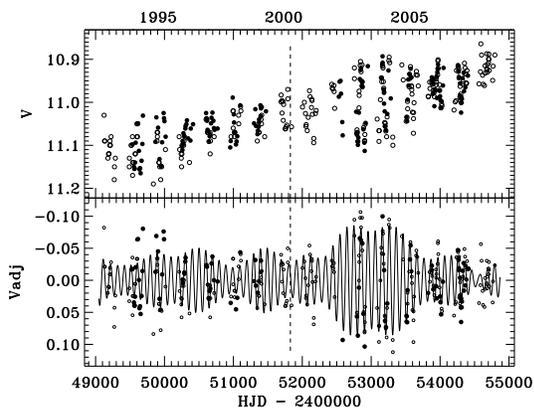}
\caption{Top: Combined {\it V} light curve of IRAS 19386+0155, with our data shown as filled circles and the data of Arkhipova et al. shown as open circles.
Bottom: The trend-removed, seasonally-adjusted {\it V} light curve fitted with the five periods and amplitudes listed in Table~\ref{lc_per}.  The fit is reasonably good over this 16-year interval.
The vertical dashed line shows the time when published high-resolution spectra were taken.
\label{19386_fit}}\epsscale{1.0}
\end{figure}

A combined {\it V} light curve was thus formed by adding the Arkhipova et al. data from 1993$-$2008 with ours, with the offset included.  This is shown in Figure~\ref{19386_fit}.  
The data sets show good agreement, both in the overall trend in brightness and in the seasonal cyclical variations present.

\placefigure{19386_fit}

\subsubsection{Period analyses}

The combined VUO and Arkhipova et al. {\it V} light curve of IRAS 19386+0155 was analyzed for variability.  The trend of increasing brightness was first removed by fitting it with a low-order polynomial.
The light curve analysis shows two dominant periods in the data, with P$_1$ = 101.8 $\pm$0.1 days and P$_2$ = 98.6 $\pm$0.1 days.  The frequency spectrum and phased light curve for P$_1$ are shown in Figure~\ref{freqspec}.  These are the same periods found by \citet{ark10} based on their data set alone.  Three additional significant periods are found in the combined data, and these are listed in Table~\ref{lc_per}.  
Similar results were found, whether or not we adjust the data to the means of the individual seasons.  We have listed in the table and shown in the figures the results with data adjusted (set A).  The fit is reasonably good over this 16-year interval.
Based on the combined, full {\it V} light curve (set A), we find a ratio of P$_2$/P$_1$ = 0.97.
We then investigated the periodicities using subsets of the adjusted {\it V} data $-$ from 
1993$-$2000 (set B) and 2001$-$2008 (set C) $-$ and also the VUO {\it R$_C$} data from 2002$-$2007.  
They result in dominant periods ranging from 99 to 103 days, similar to those found in the combined, full {\it V} data set.
We also investigated the ASAS {\it V} data set from 2002$-$2009, which yielded a dominant period of 102.3 $\pm$ 0.1 days, with secondary periods of  83 and 101 days. The 2002$-$2005 ASAS data, which appear the most cyclical, yield a dominant period of 102.8 $\pm$ 0.2 days.
These ASAS results agree with both the period value cited by \citet{ark10} of 102$\pm$1 day based upon the 2003$-$2004 ASAS data alone, and
they agree with the values found from our larger and more precise data sets.

\subsection{IRAS 19475+3119}
\subsubsection{Combined light curves}

IRAS 19475+3119 was observed by \citet{ark06} over a four-year interval from 2002 to 2005, resulting in 104 {\it UBV} measurements.  They reported that the object showed semi-regular brightness variations with a maximum range of 0.17 mag in {\it V}, and a larger range in the shorter bandpasses.  This agrees with the variations that we find.  They also reported that they did not find a statistically-significant period for the entire data set, but periods were found for individual years of 43$\pm$1 day for 2003 and 45$\pm$1 day for 2004.  
Upon downloading their archived data, 
we found that it included some more observations from the following season, for a total of 132 {\it UBV} measurements.

\begin{figure}\figurenum{11}\epsscale{0.85}
\plotone{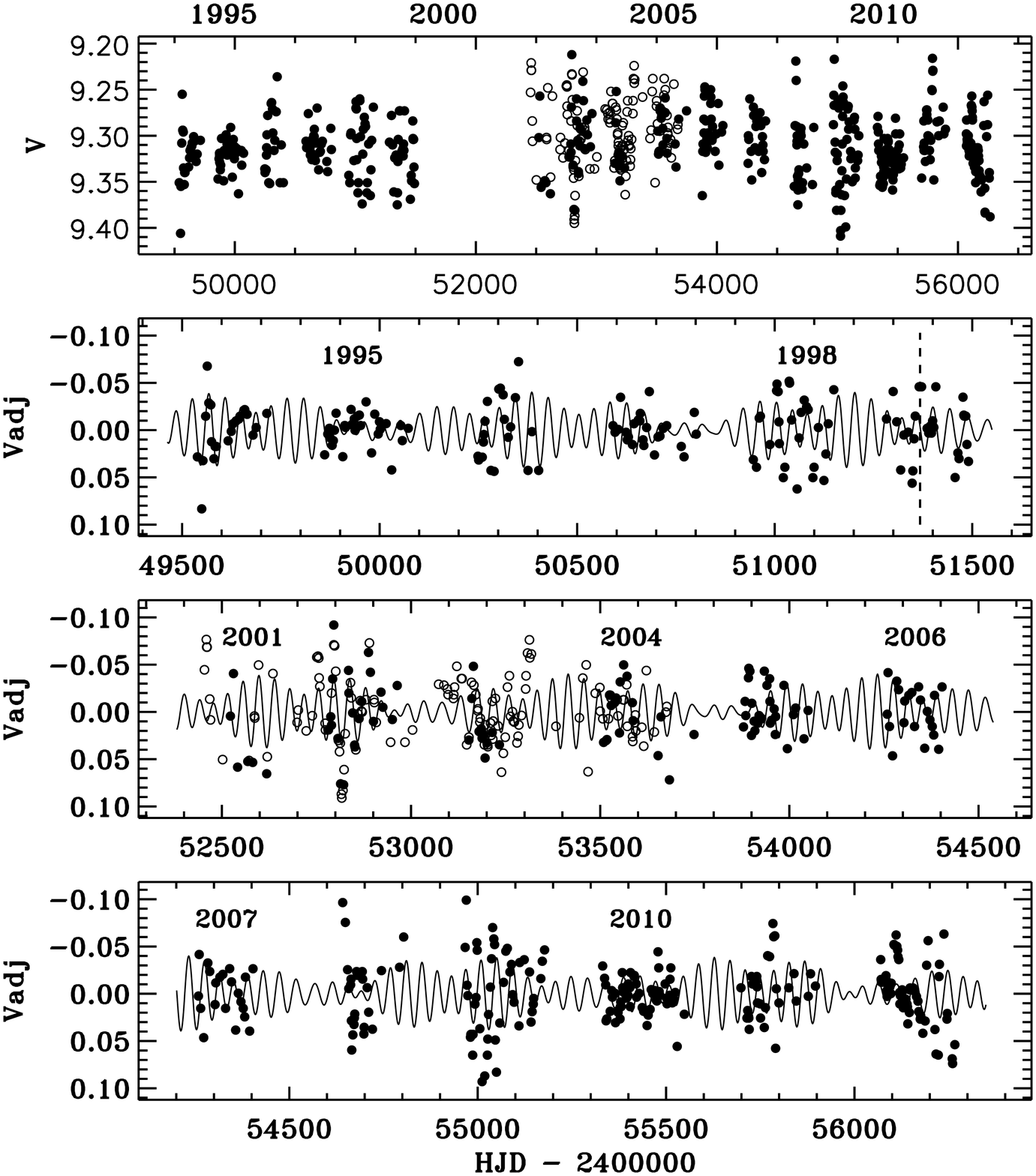}
\caption{Top: Combined {\it V} light curve of IRAS 19475+3119, with our data shown as filled circles and the data of Arkhipova et al. shown as open circles.
Bottom three panels: The seasonally-adjusted {\it V} light curve fitted with the three periods and amplitudes listed in Table~\ref{lc_per}.  The fit is not so good, particularly in the amplitudes.  We attribute this to the complex nature of the light curves, likely due to the presence of shock waves in the atmosphere.
 Note that there is some repetition of the data at the edges of the two lower panels.
 The vertical dashed line shows the time when published high-resolution spectra were taken.
\label{19475_fit}}\epsscale{1.0}
\end{figure}

Since their data set falls within the time range of ours, including their data will not extend the base line but will increase the density of observations during that five-year interval of overlap.   
Comparing {\it V} observations from the two data sets revealed an offset in observations made on the same day of (VUO$-$Arkhipova) of $-$0.094 mag (6 nights) and an offset of observations made on successive days of $-$0.087 mag (12 nights).  This yields a weighted offset of $-$0.091 mag, where those made on successive dates received only half weight.  
The combined {\it V} light curve with the offset included is displayed in Figure~\ref{19475_fit}.  
Visual inspection season by season shows good agreement between the two data sets.

\placefigure{19475_fit}

\subsubsection{Period analyses}

A period analysis of IRAS 19475+3119 was first carried out on the combined {\it V} light curve.
We analyzed the light curve both with and without adjusting the data to the seasonal means.  The same results were obtained in both cases, with two significant periods, P$_1$ = 40.95 $\pm$0.01 days and P$_2$ = 38.85 $\pm$0.01 days (set A).
An analysis of the normalized VUO {\it R$_C$} light curve (set B) also resulted in two similar periods.
These period results are listed in Table~\ref{lc_per}.
The frequency spectrum and the phased light curve plot based on P$_1$ for the combined and adjusted {\it V} data are shown in Figure~\ref{freqspec}.
Both the {\it V} and {\it R} light curves also gave evidence of a third period, P$_3$ = 48.1 days, with a S/N = 3.9, slightly below our significance level of 4.0.  However, given that it is seen in both of the data sets, we regard it as significant.
These each yield a ratio of P$_2$/P$_1$ = 0.95.

In Figure~\ref{19475_fit} is displayed the adjusted {\it V} light curve for the combined data, fitted with the three periods and associated amplitudes and phases.  
The fit for this star is not as good as that found for the other three, and we attribute much of this to the complexity of the pulsations of this star.
A better fit could likely be obtained with time-dependant amplitudes.  PPNs have been shown to produce shocks as they pulsate \citep{leb96,zacs09}.
A spectroscopic and photometric study of the similar (F5~I, P$\sim$39 days) but carbon-rich PPN IRAS 07134+1005 (HD 56126, CY CMi), with a similarly complex light curve, displayed strong evidence for shock waves in its atmosphere \citep{bar00}.

We also analyzed the light curves in subsets and using the other filters.
Analyses of the {\it V} (set C) and  {\it R$_C$} (set D) light curves from 2002$-$2012 yield similar periods P$_1$ and P$_2$ to the full data sets.
Only the {\it V} light curve for the interval 1994$-$1999 (set E), based on many fewer points, gave significantly different periods of 36.7 and 36.1 days, values not found in the other data sets.  This suggests a change in the period from the earlier to later time intervals.  However, this result is based on a smaller data set over fewer years, and thus these periods are less secure than the periods found in the other data sets for this object.

\section{DISCUSSION}
\label{disc}

We have analyzed light curves that span time intervals from 16 to 34 years for four bright, O-rich PPNs, all with F spectral types, 
Dominant periods were determined for all four, with values ranging from 41 to 113 days.
Similar, although less robust values, had been found in previous studies.
These values fall within the range found for other F$-$G spectral type PPNs, in particular the 12 C-rich ones analyzed by \citet{hri10} that span the period range of 38 to 153 days.  A similar range was found for eight C-rich PPNs in the Magellanic Clouds \citep[Paper III]{hri15a}.

We also analyzed the light curves within subsets of five
to ten year time intervals.  
These analyses often resulted in different dominant period values in different time intervals, indicating that significant period changes do occur.  
In a majority of cases, P$_1$ increased with time, but not in all cases.  
In a previous period analysis of two C-rich PPNs with similarly long sets of observations, we found that in one case there was no significant change in P$_1$ and in the other P$_1$ decreased \citep{hri13}.
With this present small sample, we conclude that no clear trends in period change emerge when comparing the data sets for an individual star in different time intervals.
The results of the larger sample of 12 PPNs had suggested that P$_1$ decreases with time \citep{hri10}, but that result is not seen over the longer two-decade time span of these studies.

Multiple periods were found for all four PPNs, and they were always close in value to the dominant period.
Similar results have been found in detailed light curve studies of other PPNs \citep{hri13,hri15a}.
The ratios of P$_2$/P$_1$ for these four range from 0.86 to 1.05, with an average value of 0.96.  
These are summarized in Table~\ref{results}.
They are similar to the ratios found in previous studies by \citet[Paper II]{hri13} for two PPNs (0.95), by \citet[Paper III]{hri15a} for seven PPNs (0.86$-$1.16), and by  \citet{ark10,ark11} for several PPNs and other post-AGB stars (0.93$-$1.06).  Also, similar values were found for other post-AGB objects by \citet{kiss07} for three objects (0.90$-$0.94) and by \citet{vanwin09} for two others  (0.91, 0.96).
Thus these seem to be typical period ratios for these low- and intermediate-mass post-AGB objects.
They presumably relate to the pulsational modes excited, although these have not been identified.

\placetable{results}

Pulsation periods and amplitudes, together with the determined temperatures, can potentially be used with stellar pulsational models to determine the masses and luminosities of these evolved objects \citep{fok01,aik10}.
These derived values can be compared to theoretical mass$-$luminosity relationships derived for post-AGB stars \citep{blo95,vas94}.  
At present, there are only a few investigations of pulsation of post-AGB stars, and the agreement with observations is not good.
\citet{aik10} ran linear, radial pulsation models of post-AGB stars with masses of 0.6 and 0.8 M$_{\Sun}$ and presented graphically the results for the temperature range 5000$-$7000 K and log~{\it g} in the range of 0.0$-$2.0.
We examined the results for T$_{\rm eff}$ = 6300 K and T$_{\rm eff}$ = 7000 K and log~${\it g}$ in the range of 0.2 to 1.0, the closest parameters to our stars.  
The fundamental modes were all stable except for those with T$_{\rm eff}$ = 6300 K and log~${\it g}$ $\le$ 0.4.  \citet{aik10} also carried out on-linear analyses of the unstable modes.
However, the calculated values had periods that were too short (days to a few weeks) and amplitudes that were too small by a factor of three when compared to our observed results.  The models by \citet{fok01} were non-linear models of masses of 0.6 and 0.8 M$_{\Sun}$ but of temperatures in the range of 5600 to 6000 K, cooler than our stars.  We previously compared them with our observational results of two cooler PPNs, but they also resulted in periods that were too short and pulsational amplitudes that were too small \citep{hri13}.
This commends further pulsational modeling of post-AGB stars in the temperature range of 6000$-$8000 K, .

\begin{figure}\figurenum{12}\epsscale{0.65} 
\plotone{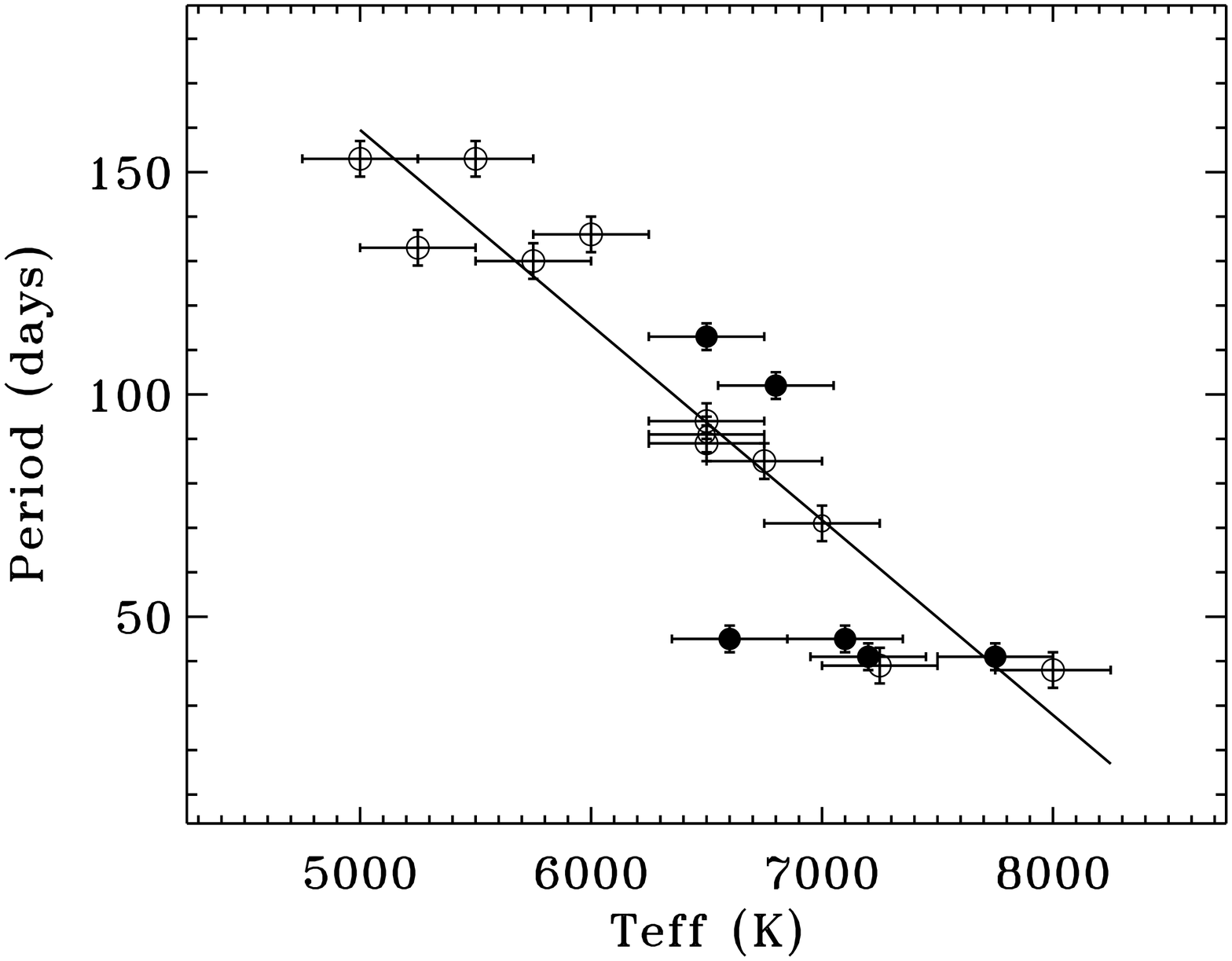}
\caption{A plot of pulsation period versus T$_{\rm eff}$.  The 12 C-rich objects are shown with open circles \citep{hri10} and the four O-rich objects from this study are shown with filled circles. 
Those with less certain values are shown with smaller symbols.   The straight line is a fit to the 12 C-rich objects.
Two of the O-rich objects (IRAS 17436+5003 and 19475+3119) have two points each because they have two different temperature measurements. 
(We have assumed reasonable error bars of $\pm$250 K and $\pm$3 days.) 
\label{P-T}}
\epsscale{1.0}
\end{figure}

\begin{figure}\figurenum{13}\epsscale{0.65} 
\plotone{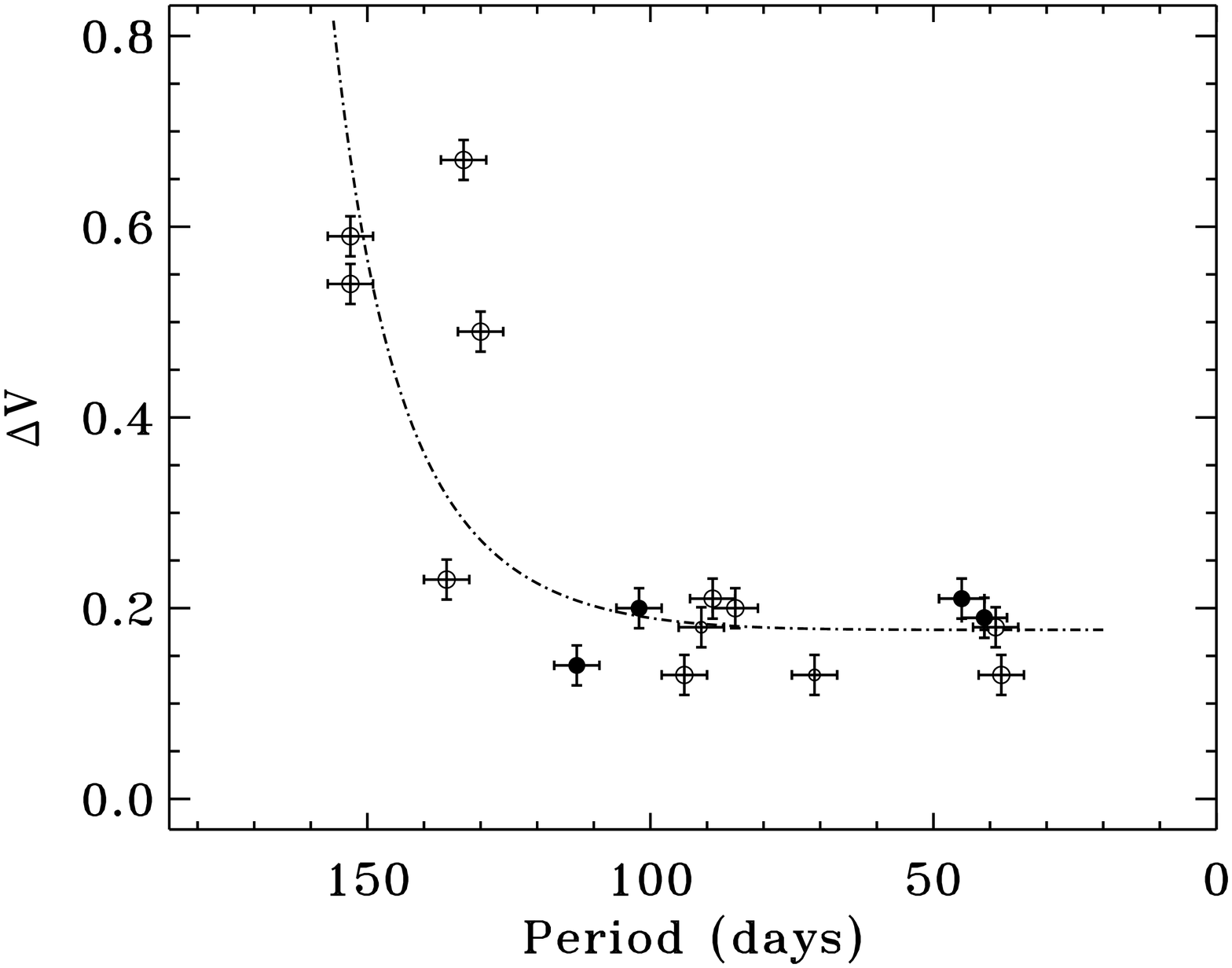}
\plotone{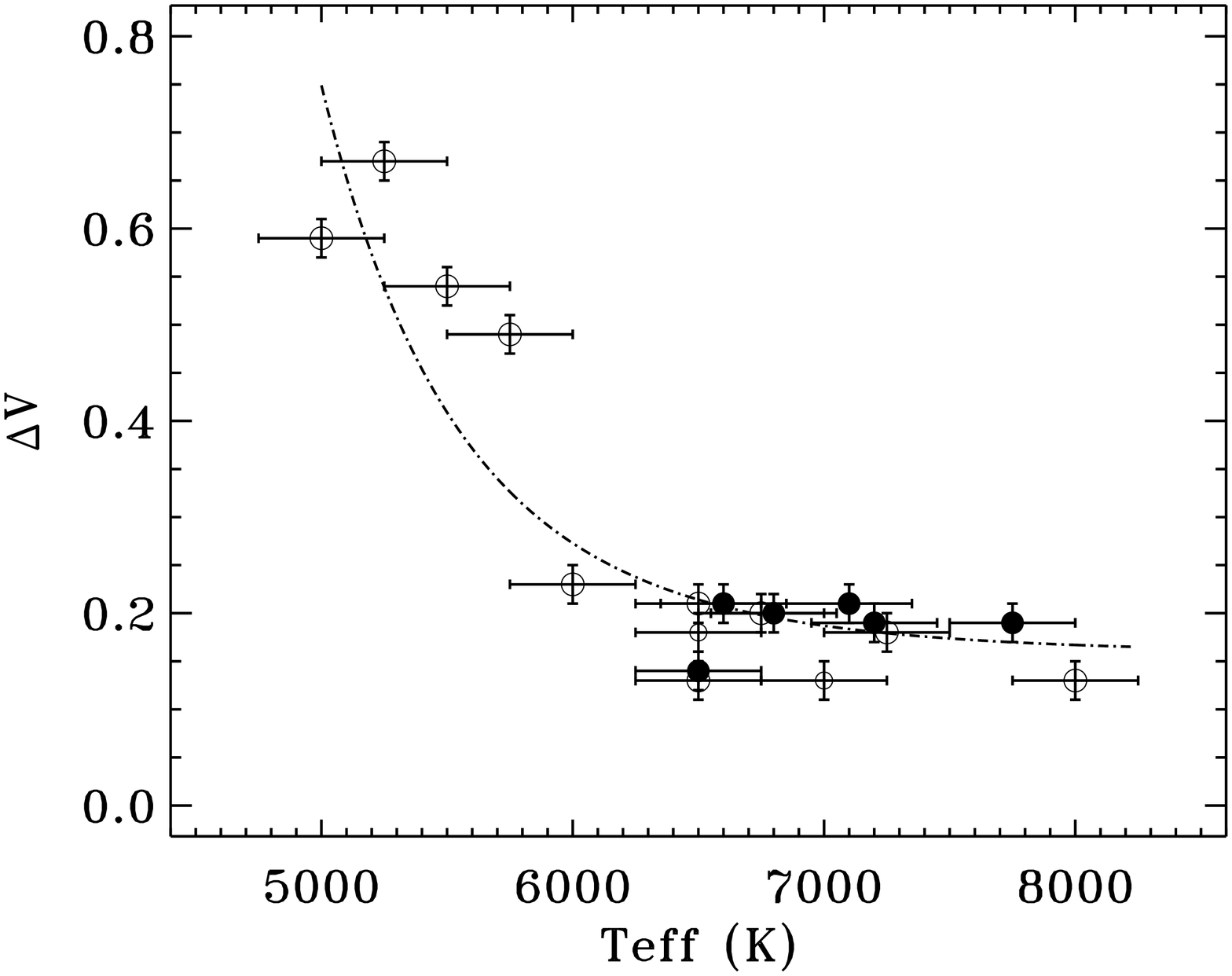}
\caption{Plots showing the comparison of the trends in the values of the maximum amplitude ($\Delta$V) in a season with period (top) and temperature (bottom).  The open circles show the values for the 12 C-rich PPNs and the dashed lines show the general trends \citep{hri10}.  Those with less certain values are shown with smaller symbols.  The filled circles show the values for these four O-rich PPNs.  The agreement is good between the two sets of data: they all show low amplitudes for shorter periods and higher temperatures.  (Error bars error bars of $\pm$250 K, $\pm$3 days, and $\pm$0.02 mag.)
\label{PT-amp}}
\epsscale{1.0}
\end{figure}

One can also make other comparisons with the results of our study of 12 C-rich PPNs in the Milky Way Galaxy.
For that sample of F$-$G stars, we \citep{hri10} found a monotonic, approximately linear decrease in period ({\it P}) with effective temperature ({\it T$_{eff}$}) over the range of 5000 to 8000 K.  
While we do not have a long enough observing interval to see an individual star change in {\it T$_{eff}$} or {\it P}, we can interpret this trend of decreasing {\it P} with increasing {\it T$_{eff}$} to represent the average expected evolution of these stars as they evolve from large, cool AGB stars to the small, hot central stars of PNs.
Since they evolve toward higher temperatures at approximately constant luminosity, they must be decreasing in size and increasing in density (ongoing mass loss is small).  
Therefore, assuming radial pulsations, we would expect a decrease in period as they evolve.
A comparison with five C-rich PPNs in the Large Magellanic Cloud (LMC) showed that they also follow a similar monotonic trend, although perhaps offset slightly to lower {\it T$_{eff}$} or shorter {\it P} (observations of more periodic LMC PPNs are needed to confirm this offset).
In Figure~\ref{P-T} is plotted the {\it P} and {\it T$_{eff}$} values for the four O-rich PPNs examined in this study, together with the 12 C-rich ones from the Milky Way Galaxy.
Note that we have included two points each for IRAS 17436+5003 and IRAS 19475+3119, since there are two different {\it T$_{eff}$} values published for each.  
A straight line fit to the data points for these 4 O-rich PPNs would be nearly vertical and possess a much steeper than that found for the C-rich ones.  However, this present sample is small and all of the points, except one of those for IRAS 17436+5003, fall within the range of the C-rich data points.
When we compare the pulsation amplitude ($\Delta$V) with their periods and temperatures (see Table~\ref{results}), they agree well with the monotonic trends found for the C-rich sample; all of the objects with periods shorter than 120 days and temperatures higher than 6000 K have maximum amplitudes $\le$0.22 mag.  This is shown in Figure~\ref{PT-amp}.

\placefigure{P-T}

\placefigure{PT-amp}

As noted above, the {\it P}$-${\it T$_{eff}$} relationship for these four O-rich PPNs appears to have a much steeper slope than is found for the 12 C-rich PPNs.  This may not be significant, since it is based on only four objects over a smaller temperature range (1000 K), and the O-rich PPNs generally fall within the distribution of the C-rich sample.  
If it is real, one suggestion is that it relates to the difference in metallicity between the two longer-period objects (P$\approx$110 day, [Fe/H]$\approx$$-$1.0) and the two shorter-period objects  (P$\approx$45 day, [Fe/H]$\approx$$-$0.25).
This would indicate that the longer-period ones derive from a more metal-poor population and thus are older and of lower mass.
The non-linear radiative pulsation models by \citet{fok01} indicate longer periods for lower-mass stars, which is consistent with what we find here.  However, this is only suggestive and may not be relevant since the models were for somewhat cooler (5600$-$6000 K) stars than these four.
If this difference in slope is confirmed by a larger study, then further theoretical models to investigate this would be desirable.

Two of the four objects in this sample show significant long-term trends in brightness over the observing interval, both getting brighter: IRAS 18095+2704 by 0.30 mag over 19 years or 0.016 mag yr$^{-1}$ and IRAS 19386+0155 by 0.13 mag over 14 years or 0.009 mag yr$^{-1}$.
Similar monotonic trends, some decreasing and some increasing, have been found in the light curves of other PPNs of F$-$G spectral types. 
Including these four, periods have been determined for a total of 26 PPNs of F$-$G spectral types, 19 in the MWG and 7 in the Magellanic Clouds, for which observations cover intervals of at least nine years 
\citep{hri10, ark10, ark11, hri15a}.
Of these, five show monotonic increases of $\sim$0.01$-$0.02 mag yr$^{-1}$,
two show monotonic decreases of $\sim$0.01 mag yr$^{-1}$, and
one shows a light curve that is constant in brightness for six years, then shows a sudden drop of 0.12 mag in one year, and then a gradual increase of 0.019 mag over nine years.
Two others show some larger changes ($\sim$0.1 mag) that are not monotonic.
The majority (16 of 26) do not show much change in the seasonal brightness levels.
  
Gradual changes in the brightness could be due to secular changes in the dust opacity.  
PPNs are surrounded by an expanding envelope of circumstellar gas and dust, which formed from mass loss on the AGB.  This circumstellar envelope can both dim and redden the emerging light. 
The sudden drop and reddening witnessed in one PPN (IRAS 19500$-$1709) has been attributed to the formation or transit of a circumstellar dust cloud \citep{hri10,ark10}.  
Both IRAS 18095+2704 and 19386+0155 are significantly reddened \citep[see Table~\ref{object_list}; (B$-$V) = 0.23-0.32 for F2$-$5I,][]{cox00}.  
However, the lack of color change in these two PPNs when they increase in brightness would require that the dust opacity along the line of site be due to large ($\ge$ 1 $\mu$m) grains, which would appear grey and not color selective, as noted by \citet{ark10}.
 
All four of these PPNs are redder, and thus cooler, when fainter.  This is what is commonly found in PPNs of F$-$G spectral types \citep{ark10, ark11, hri10, hri15a}.  
 \citet{hri13} found in the detailed study of two PPNs that included contemporaneous light, color, and velocity curves, that the stars were smallest when brightest and hottest.  This differs from what is found in Cepheid variables, in which there is a phase lag of $\sim$0.25 between the maximum in light and the minimum of size. 
This temperature and size correlation will be investigated for three of these O-rich PPNs through a coordinated light, color, and velocity study, for which observations are presently being made.

The specific changes in color can be correlated to specific changes in temperature for these four objects.   The maximum ranges in colors are listed in Table~\ref{results}.  Using the observed ({\it B$-$V}) changes and the temperatures determined from the spectroscopic analyses, these color changes were transformed to temperature changes using the color-temperature table of  \citet[Table 15.7]{cox00}.  This resulted in maximum temperature changes of 400 to 770 K, which are attributed to pulsation.  
Of the two PPNs with multiple spectroscopically-determined temperature values, for one of them (IRAS 17436+5003) the difference between the two spectroscopic temperatures is less than the range determined from the color range, and for one of them (IRAS 19475+3119), the difference is slightly greater, but within the uncertainty of the temperature determinations.
We investigated the epochs of the high-resolution spectra used to determine the temperatures, to see if they could be correlated with specific phases in the light curves.  For IRAS 17436+5003, the higher temperature spectrum was obtained near maximum brightness and bluest color (see Fig.~\ref{17436_fit}), while the lower temperature spectrum was obtained when there were no contemporaneous light curves.  For IRAS 18095+2704, the spectrum was taken near maximum brightness and bluest color\footnote{\citet{sah11} obtained two separate spectra, observed 10 months apart.  Since they based the analysis primarily on the first of these, this is the epoch we investigated. } (see Fig.~\ref{18095_fit}), and for IRAS 19386+0155, between maximum and average brightness (see Fig.~\ref{19386_fit}).  Of the two spectra of IRAS 19475+3119, the higher temperature one was obtained near maximum brightness and the lower temperature one at an epoch with few photometric observations and no clear phasing of the brightness.  Thus, for the two objects with multiple spectroscopic temperature determinations, the higher temperatures are associated with the greater brightness in the light curves, consistent with the brightness-color correlations found from the light curves.

\section{SUMMARY AND CONCLUSIONS}
\label{summary}

In this paper, we have carried out a detailed light curve and period study of four bright PPNs. They each had previous light curve and period studies, but by combining these data sets with ours, we have significantly increased the sample size and the time intervals of observation.  The primary results are listed below.  They serve to help elucidate the pulsation properties of intermediate- and low-mass post-AGB during this short transition stage from the AGB to the PN phases.

1. The four vary in light with changing amplitudes; these amplitudes vary between 0.05 and 0.25 mag, peak to peak.

2. Two of them also show long-term trends of increasing brightness, with values of 0.13 and 0.30 mag over 14 and 19 years, respectively, corresponding to 0.01$-$0.02 mag yr$^{-1}$. 
Long-term trends have been observed in about 30$\%$ of the well-studied PPNs, with five showing clear monotonic trends of increased brightness and two of decreased brightness.  We attribute these to changes in the circumstellar dust opacity.

3. A dominant period was found in each of the four, with values ranging from 41 to 113 days.  Similar periods had previously been found for three of these based in smaller data sets.
These values fall in the period range found previously for PPNs of F$-$G spectral types of 38 to 153 days.

4. In each of the four PPNs, multiple periods were determined.  These account for much of the variation in amplitude seen in the light curves.  In all cases, the secondary period is close to the primary period, with an average value of P$_2$/P$_1$ of 0.96.  This is similar to the range of values of 0.9$-$1.1 found in previously studied PPNs with suitably long temporal data sets.  These multiple periods help to explain the changing light curve amplitudes as they beat together.  They also help to explain the seasonally changing periods attributed to IRAS 17436, which instead are due to multiple periods affecting the light curve shape.

5. For these four, one finds a general trend of the cooler ones having longer periods.  However, this is not a strong trend and the sample is small.  All four of these are O-rich.  A strong trend has been seen in C-rich F$-$G spectral type PPNs. These four O-rich ones have a steeper slope to their trend, but the sample is too small to determine if the slopes differ between PPNs with different chemistries.

6. All four are redder when fainter, in agreement with the brightness-color relationship found in other well-studied PPNs.  The maximum (B$-$V) color changes range from 0.055 to 0.105 mag, which imply changes in effective temperature of 400$-$770 K.

We have begun photometric studies of additional, fainter O-rich PPNs, which will help to enlarge the sample and increase the range in temperature.  These should help to better define the period-temperature-amplitude relationships for O-rich PPNs and allow a more robust comparison with C-rich PPNs.

\acknowledgments We want to acknowledge the many VU undergraduate summer research students who participated in this long-term research program: 
Danielle Boyd, Laura Nickerson, Paul Barajas, Jason Webb, Bradley Spitzbart, 
George Lessmann, Will Herron, Richard Maupin, Emily Cronin, Ryan Doering, Shannon Pankratz, Andrew Juell, Daniel Allen, Justin Lowry, Kathy Cooksey, Jeffrey Eaton, Nicolas George, Katie Musick, Sarah Schlobohm, Brian Bucher, Kara Klinke, Kristina Wehmeyer, Bradley Rush, Byung-Hoon Yoon, Jeffrey Massura, Marta Stoeckel, Larry Selvey, Jason Strains, Ansel Hillmer, Erin Lueck, Joseph Malan, Callista Steele, Ryan McGuire, Christopher Wagner, Samuel Schaub,  Zachary Nault, 
Wesley Cheek, Joel Rogers, Rachael Jensema, Christopher Miko, Austin Bain, Hannah Rotter, and Aaron Seider.
The ongoing work of Paul Nord in maintaining the equipment is gratefully acknowledged.
We thank Kevin Volk for ongoing conversations about evolved stars.
We also thank the anonymous referee for her/his suggestions which improved the presentation
of these results.
Equipment support for the VU Observatory was partially provided by grants from the
National Science Foundation College Science Instrumentation Program (8750722), the Lilly ``Dream of Distinction'' Program, and the Juenemann Foundation.  BJH acknowledges onging support during this project from the National Science Foundation (9315107, 9900846, 0407087, 1009974, 1413660), NASA through the JOVE program, and the Indiana Space Grant Consortium.
This research has made use of the SIMBAD database, operated at CDS, Strasbourg,
France, and NASA's Astrophysical Data System.

\clearpage

\tablenum{1}
\begin{deluxetable}{crclrrrccl}
\tablecaption{Program Objects\label{object_list}}
\tabletypesize{\footnotesize} 
\tablewidth{0pt} \tablehead{
\colhead{IRAS ID}&\colhead{V\tablenotemark{a}}
&\colhead{(V$-$R$_C$)\tablenotemark{a,b}}&\colhead{Sp.T.}&\colhead{T$_{\rm eff}$}
&\colhead{log {\it g}}&\colhead{[Fe/H]}
&\colhead{[C/O]}&\colhead{Ref.\tablenotemark{c}}&\colhead{Other ID}\\
\colhead{}&\colhead{(mag)}&\colhead{(mag)}&\colhead{}
&\colhead{(K)}&\colhead{}&\colhead{}&\colhead{}&\colhead{}&\colhead{}} \startdata
17436+5003 & 7.0 & 0.2 & F3 Ib\tablenotemark{d} & 6600 & 0.0 & $-$0.3 & $-$0.2 &1 & HD 161796, V814 Her \\
                   & 7.0 & 0.2 & F3 Ib\tablenotemark{d} & 7100 & 0.5 & $-$0.25 & $-$0.3 &2 & \nodata \\
18095+2704 & 10.0 & 0.7 & F3 Ib & 6500 & 0.5 & $-$0.9 & $-$0.37 &3 &V887 Her \\
19386+0155 & 11.1 & 0.8 & F5 Ib & 6800 & 1.4 & $-$1.1 & $-$0.35 & 4 & V1648 Aql \\
19475+3119 &  9.3 & 0.4 & F3 Ib & 7750 & 1.0 & $-$0.24 & $-$0.39 &5 & HD 331319, V2513 Cyg  \\
                   & 9.3 & 0.4 & F3 Ib & 7200 & 0.5 & $-$0.25 & $-$0.5 &2 & \nodata \\
\enddata
\tablenotetext{a}{Variable. }
\tablenotetext{b}{Includes circumstellar and interstellar reddening. }
\tablenotetext{c}{References for the spectroscopic analyses: (1) \citet{luck90}, (2) \citet{kloch02}, (3) \citet{sah11}, (4) \citet{per04}. (5) \citet{ferro01}.  Note that there is an earlier spectroscopic study of IRAS 18095+2704 by \citet{kloch95}, but the one cited is at $\sim$2.5 times higher resolution. } 
\tablenotetext{d}{Classified as A7~I by \citet{sua06}. } 
\end{deluxetable}

\clearpage

\tablenum{2}
\begin{deluxetable}{llrrrcr}
\tablecaption{Standard Magnitudes of Program and Comparison
Stars\tablenotemark{a} \label{phot_std}} 
\tabletypesize{\footnotesize} 
\tablewidth{0pt} \tablehead{
\colhead{Object} &\colhead{GSC ID} &\colhead{V} &\colhead{(B$-$V)} &\colhead{(V$-$R$_C$)}
&\colhead{(R$_C$$-$I$_C$)} & Run\tablenotemark{b}}
\startdata
IRAS~17436+5003 & 03518-00402 &   7.00 & \nodata  & 0.25 & \nodata & 1 \\
				&   & 6.94 & 0.35  & 0.22 & \nodata & 2 \\
C$_1$\tablenotemark{c} & 03518-00926 & 10.32 & 0.81 & 0.46 & \nodata & \nodata\\
C$_2$                    	& 03518-01037\tablenotemark{d} &  9.65 & 1.05 & 0.58 & \nodata & \nodata \\
C$_3$                    	& 03518-01174\tablenotemark{e} &  9.64 & 1.03 & 0.59 & \nodata & \nodata \\
\\
IRAS~18095+2704  & 02100-00044 & 10.30 & \nodata & 0.72 & \nodata & 1\\
				&  & 10.00 & 1.02 & 0.69 & \nodata & 2\\
C$_1$                      & 02100-00387\tablenotemark{f} & 12.28 & 0.55 & 0.33 & \nodata & \nodata \\
C$_2$\tablenotemark{g}  & 02100-00238\tablenotemark{h} & 12.41 & 0.88 & 0.54 & \nodata & \nodata \\
C$_3$                      & 02100-00003 & 13.57 & 0.98 & 0.56 & \nodata & \nodata \\
\\
IRAS~19386+0155 & 00483-00956 &  11.09 & 1.16 & 0.77 & 0.86 & 3\\
C$_1$                    & 00483-01010 & 13.04 & 0.81 & 0.46 & 0.49 &  \nodata\\
C$_2$                    & 00483-01014 & 13.44 & 0.94\tablenotemark{i} & 0.55 & 0.56 & \nodata\\
C$_3$                    & 00483-01082 & 13.54 & 0.83\tablenotemark{i} & 0.46\tablenotemark{i} & 0.51\tablenotemark{i} & \nodata \\
\\
IRAS~19475+3119  & 02669-01757 &  9.27 & \nodata & 0.41 & \nodata & 1\\
				&  &  9.32 & 0.63 & 0.40 & \nodata & 2\\
C$_1$\tablenotemark{j}  & 02669-02709 &11.02 & 0.14 & 0.08 & \nodata & \nodata \\
C$_2$                     & 02669-04063 &11.92 & 1.76 & 1.01 & \nodata & \nodata \\
C$_3$\tablenotemark{k} & 02669-04215 &12.39 & 1.32 & 0.75 & \nodata & \nodata \\
\enddata
\tablenotetext{a}{Uncertainties in the observations are as follows $-$ {\it V}: $\pm$0.01, 
({\it B$-$V}): $\pm$0.025, ({\it V$-$R$_C$}): $\pm$0.015 mag, and ({\it R$_C$$-$I$_C$}): $\pm$0.010 mag, except as noted. }
\tablenotetext{b}{Standardized observations were made at the VUO on UT (1) 22 August 1995 and (2) 26 June 2012, and are averaged together for the comparison stars, 
and at (3) the Kitt Peak National Observatory on 23 June 1994. }
\tablenotetext{c}{Evidence of a general increase in brightness from 1994$-$2007 of 0.02 mag ({\it V}).}
\tablenotetext{d}{HD 234482}
\tablenotetext{e}{HD 234480}
\tablenotetext{f}{TYC 2100-387-1}
\tablenotetext{g}{Suggestion of a gradual monotonic increase in brightness of 0.025 mag ({\it V}) from 1994 to 2007 
and a gradual increase and then decrease over a range of 0.03 mag ({\it V}) from 2008 to 2012.  }
\tablenotetext{h}{TYC 2100-238-1}
\tablenotetext{i}{Uncertainty of $\pm$0.02 mag. }
\tablenotetext{j}{Suggestion of variation $\le$ 0.03 mag peak to peak. }
\tablenotetext{k}{Suggestion of long-term ($\ge$ 8 year) cyclical variation $\sim$0.03 mag peak to peak. }
\end{deluxetable}

\clearpage

\tablenum{3}
\begin{deluxetable}{rrrrrcrrrrcr}
\rotate \tablenum{3} 
\tablecolumns{12} \tabletypesize{\scriptsize}
\tablecaption{Statistics of the VUO PPN Light and Color Curves\label{tab_stat}}
\tabletypesize{\footnotesize}
\tablewidth{0pt} \tablehead{ 
\colhead{IRAS ID} & \colhead{Years} && \multicolumn{4}{c}{Number of Observations}
&& \multicolumn{4}{c}{Average Uncertainty (mag)\tablenotemark{a}} \\
\cline{4-7} \cline{9-12} 
\colhead{} & \colhead{}&&
\colhead{$\Delta$V}&\colhead{$\Delta$R$_C$} &\colhead{$\Delta$(V$-$R$_C$)} & \colhead{$\Delta$B} &&
\colhead{$\sigma$($\Delta$V)} &\colhead{$\sigma$($\Delta$R$_C$)} &\colhead{$\sigma$($\Delta$(V-R$_C$))} & \colhead{$\sigma$($\Delta$B)}
} \startdata
17436$+$5003 & 1994$-$2012 && 563 & 542 & 450 & 284 && 0.006 & 0.006 & 0.009 & 0.008 \\
18095$+$2704 & 1994$-$2012 && 530 & 473 & 399 & 246 && 0.005 & 0.005 & 0.007 & 0.005 \\
19386$+$0155 & 1994$-$2007 && 177 & 135 & 109 & \nodata && 0.009 & 0.006 & 0.010 & \nodata\\
19475$+$3119 & 1994$-$2012 && 481 & 425 & 360 & 236 && 0.003 & 0.004 & 0.005 & 0.002\\
\enddata
\tablenotetext{a}{The average statistical uncertainty in a single differential measurement.}
\end{deluxetable}

\clearpage

\tablenum{4}
\begin{deluxetable}{lrrrrrr}
\tablecolumns{7} \tabletypesize{\scriptsize}
\tablecaption{Differential Standard {\it VR$_C$} Magnitudes and Colors for
the Four PPNs from 1994$-$2007\tablenotemark{a,b} \label{diffmagVR_std}}
\tabletypesize{\footnotesize}
\tablewidth{0pt} \tablehead{ \colhead{Program Star} & \colhead{HJD $-$ 2,400,000}
&\colhead{$\Delta$V} 
&\colhead{HJD $-$ 2,400,000} &\colhead{$\Delta$R$_C$}
 &\colhead{HJD $-$ 2,400,000}
&\colhead{$\Delta$(V$-$R$_C$)} }
\startdata
IRAS 17436+5003 & 49557.6197 & $-$3.334 & 49873.6310 & $-$3.121 & 49873.6309 & $-$0.222 \\
IRAS 17436+5003 & 49565.5887 & $-$3.311 & 49882.6152 & $-$3.085 & 49882.6117 & $-$0.200 \\
IRAS 17436+5003 & 49571.5873 & $-$3.268 & 49943.6121 & $-$3.092 & 49943.6121 & $-$0.245 \\
IRAS 17436+5003 & 49615.5535 & $-$3.302 & 49951.5862 & $-$3.115 & 49951.5869 & $-$0.221 \\
IRAS 17436+5003 & 49624.5247 & $-$3.256 & 50263.6893 & $-$3.146 & 50263.6889 & $-$0.227 \\
IRAS 17436+5003 & 49632.5300 & $-$3.279 & 50276.6724 & $-$3.103 & 50276.6724 & $-$0.214 \\
IRAS 17436+5003 & 49638.5021 & $-$3.295 & 50288.6915 & $-$3.092 & 50288.6914 & $-$0.201 \\
IRAS 17436+5003 & 49643.5242 & $-$3.341 & 50385.4892 & $-$3.122 & 50385.4892 & $-$0.209 \\
IRAS 17436+5003 & 49646.5125 & $-$3.363 & 50591.6397 & $-$3.058 & 50591.6431 & $-$0.207 \\
IRAS 17436+5003 & 49653.4942 & $-$3.339 & 50596.7252 & $-$3.074 & 50596.7250 & $-$0.216 \\
\enddata
\tablecomments{Average uncertainties in the brightness are as follows: $\pm$0.006 ($\Delta$${\it V}$), 
$\pm$0.005 ({$\Delta$R$_C$}), $\pm$0.008 mag ($\Delta$(V$-$R$_C$)), 
with maximum uncertainties $\sim$2 times as great. }
\tablenotetext{a}{Table~\ref{diffmagVR_std} published in its entirety in the
electronic edition of the Astronomical Journal.  A portion of Table~\ref{diffmagVR_std} is
shown here for guidance regarding form and content.}
\tablenotetext{b}{The 2007 observations that contain all three {\it BVR$_C$} filters are listed instead in Table~\ref{diffmagBVR_std}. }
\end{deluxetable}

\clearpage

\tablenum{5}
\begin{deluxetable}{lcrrrcrrrr}
\tablecolumns{14} \tabletypesize{\scriptsize}
\tablecaption{Differential Standard {\it BVR$_C$} Magnitudes for the Four PPNs from 2007$-$2012\tablenotemark{a,b} 
 \label{diffmagBVR_std}}
\tablewidth{0pt} \tablehead{ \colhead{Program Star} & \colhead{HJD $-$ 2,400,000\tablenotemark{c}}
&\colhead{$\Delta$B}  &\colhead{$\Delta$V} &\colhead{$\Delta$R$_C$} & &
\colhead{HJD $-$ 2,400,000\tablenotemark{c}}
&\colhead{$\Delta$B}  &\colhead{$\Delta$V} &\colhead{$\Delta$R$_C$}}
\startdata
IRAS 17436+5003 & 54283.6210 & $-$3.761 & $-$3.304 & $-$3.098 && 55415.6360 & \nodata  & $-$3.301 &   \nodata  \\
IRAS 17436+5003 & 54287.6152 & $-$3.771 & $-$3.311 & $-$3.086 && 55417.6187 & $-$3.719 & $-$3.287 & $-$3.083 \\
IRAS 17436+5003 & 54288.6223 & $-$3.737 & $-$3.308 & $-$3.104 && 55421.5619 & $-$3.714 & $-$3.279 & $-$3.077 \\
IRAS 17436+5003 & 54301.6111 & $-$3.838 & $-$3.344 & $-$3.135 && 55428.5967 & $-$3.721 & $-$3.298 & $-$3.080 \\
IRAS 17436+5003 & 54302.6374 & $-$3.834 & $-$3.364 & $-$3.161 && 55436.6151 & $-$3.802 & $-$3.339 & $-$3.116 \\
IRAS 17436+5003 & 54304.6510 & $-$3.851 &  \nodata  & $-$3.128 && 55438.6311 & $-$3.803 & $-$3.367 & $-$3.137 \\
IRAS 17436+5003 & 54304.6523 & $-$3.851 &  \nodata & $-$3.128 && 55440.6121 & $-$3.836 & $-$3.390 &  \nodata  \\
IRAS 17436+5003 & 54305.6554 & $-$3.853 & $-$3.368 & $-$3.148 && 55447.5513 & $-$3.875 & $-$3.404 & $-$3.153 \\
IRAS 17436+5003 & 54311.6534 & $-$3.796 & $-$3.339 & $-$3.122 && 55449.6145 & $-$3.870 & $-$3.394 & $-$3.158 \\
IRAS 17436+5003 & 54312.5908 & $-$3.801 & $-$3.361 & $-$3.111 && 55452.5671 & $-$3.862 & $-$3.388 & $-$3.151 \\
\enddata
\tablecomments{Average uncertainties in the brightness are as follows: $\pm$0.005 ($\Delta$${\it B}$), 
$\pm$0.005 ($\Delta$${\it V}$), and $\pm$0.004 ({$\Delta$R$_C$}),  
with maximum uncertainties $\sim$2 times as great. The uncertainties for IRAS 17436+5003 are higher than the average and those for IRAS 19475+3119 are lower.}
\tablenotetext{a}{Table~\ref{diffmagBVR_std} is published in its entirety in the 
electronic edition of the Astronomical Journal.  A portion of Table~\ref{diffmagBVR_std} 
is shown here for guidance regarding form and content.}
\tablenotetext{b}{The 2007 observations that contain only {\it VR$_C$} filters are listed instead in Table~\ref{diffmagVR_std}. }
\tablenotetext{c}{The time represents the mid-time of the {\it V} observations.  The times for the {\it B} and {\it R$_C$} observations differ from this by approximately the following: IRAS 17436+5003:   +0.0011 and $-$0.0012 days, respectively; IRAS 18095+2704: +0.0027 and $-$0.0018 days, respectively; IRAS 19475+3119: +0.0021 and $-$0.0017 days, respectively. }
\end{deluxetable}

\clearpage

\tablenum{6}
\begin{deluxetable}{clccr}
\tablecaption{Offsets Between the Published Data Sets and VUO Data Sets\tablenotemark{a} 
\label{LC_offset}}
\tablewidth{0pt} \tablehead{
\colhead{IRAS ID} & \colhead{Source} & \colhead{Dates} & \colhead{Bandpass} & \colhead{Offset (mag)}} \startdata
17436+5003 & Fernie\tablenotemark{b} &1980$-$1998 & V & 0.000 \\
                  & Percy \& Welch\tablenotemark{c} &1979$-$1980 & V & 0.000 \\
18095+2704 & Arkhipova\tablenotemark{d} &1993$-$2008& V & $-$0.048\\
                   & Arkhipova\tablenotemark{d} &1993$-$2008& B & $-$0.004\\
19386+0155 & Arkhipova\tablenotemark{d} &1993$-$2008& V & 0.060\\
                   & ASAS\tablenotemark{e} &2002$-$2009 & V & $-$0.055\\
19475+3119 & Arkhipova\tablenotemark{f} &2002$-$2005& V & $-$0.091\\
\enddata
\tablenotetext{a}{Offset = VUO $-$ published data set.}
\tablenotetext{b}{\citet{fern83,fern86,fern89,fern90a,fern90b,fern91,fern93,fern95}.  Data from 1991 to 1998 are available at http://www.astro.utoronto.ca/DDO/research/89her.html; the other data are listed in tables in the publications.}
\tablenotetext{c}{\citet{percy81}, where the data are listed.} 
\tablenotetext{d}{\citet{ark93,ark00,ark10}; data available on-line from VizieR at http://vizier.u-strasbg.fr/viz-bin/VizieR?-source=J/PAZh/26/705 and http://vizier.u-strasbg.fr/viz-bin/VizieR?-source=J/PAZh/36/281.} 
\tablenotetext{e}{http://www.astrouw.edu.pl/asas/.} 
\tablenotetext{f}{\citet{ark06}; data available on-line from VizieR at http://vizier.u-strasbg.fr/viz-bin/VizieR?-source=J/PAZh/32/48.} 
\end{deluxetable}

\clearpage

\tablenum{7}
\begin{deluxetable}{cclrrrrrrrrrrrrrr}
\tablecolumns{17} \tabletypesize{\scriptsize}
\tablecaption{Periodogram Study of the Light Curves of Four O-Rich PPNs\tablenotemark{a}\label{lc_per}}
\rotate
\tabletypesize{\footnotesize} 
\tablewidth{0pt} \tablehead{ 
\colhead{IRAS ID} &\colhead{Set} &\colhead{Filter} & \colhead{Years}&\colhead{No.} & \colhead{P$_1$}&\colhead{A$_1$} &\colhead{P$_2$}& \colhead{A$_2$} &\colhead{P$_3$} &\colhead{A$_3$}
&\colhead{P$_4$} &\colhead{A$_4$} &\colhead{P$_5$} &\colhead{A$_5$} &\colhead{P$_6$} &\colhead{A$_6$} \\
\colhead{} & \colhead{} &\colhead{} & &\colhead{Obs.} & \colhead{(days)}&\colhead{(mag)} &\colhead{(days)}& \colhead{(mag)} &\colhead{(days)} &\colhead{(mag)}
& \colhead{(days)}&\colhead{(mag)} &\colhead{(days)}& \colhead{(mag)} &\colhead{(days)} &\colhead{(mag)} }
 \startdata
17436 & A & {\it V} & 1979-2012 & 1640 & 45.15 & 0.014 & 47.36 & 0.015 & 46.75 & 0.012 & 49.89 & 0.014 & 50.38 & 0.012 & 44.96 & 0.013 \\
17436 & B & {\it V} & 1979-1998 & 1083 & 45.08 & 0.026 & 46.54 & 0.018 & 47.45 & 0.018 & 49.58 & 0.017 & 55.47 & 0.03 & 39.62 & 0.012 \\
17436 & C & {\it V} & 1994-2012 &  557  & 47.30 & 0.020 & 49.92 & 0.017 & 42.36 & 0.017 & \nodata & \nodata & \nodata & \nodata & \nodata & \nodata \\
17436 & D & {\it R$_C$} & 1999-2012 &  534 & 47.36 & 0.014 & 49.82 & 0.012 & 42.34 & 0.012 &  \nodata &  \nodata & \nodata & \nodata & \nodata  & \nodata \\
17436 & E & {\it V} & 1994-1999 & 146 & 44.09 & 0.032 & 42.38 & 0.027 & 48.90 & 0.018 & 72.43 & 0.017 & 53.89 & 0.014 & \nodata & \nodata \\
17436 &  F& {\it V} & 2003-2007 &  143  & 41.47 & 0.020 & 56.87 & 0.021 & 43.88 & 0.013 & 54.15 & 0.012 & \nodata & \nodata & \nodata & \nodata \\
17436 & G & {\it V} & 2008-2012 &  268  & 47.42 & 0.030 & 57.70 & 0.022 & 38.14 & 0.019 & \nodata & \nodata & \nodata & \nodata & \nodata & \nodata \\
\\
18095 & A & {\it V} & 1993-2012 & 844 & 113.3 & 0.017 & 97.9 & 0.011 & 103.0 & 0.010 & 108.4 & 0.009 & 158.9 & 0.009 & 116.1 & 0.009 \\
18095 & B & {\it V} & 1993-2002\tablenotemark{b} & 327 &  98.2 & 0.015 & 107.5 & 0.013 & \nodata & \nodata & \nodata & \nodata & \nodata & \nodata & \nodata & \nodata \\
18095 & C & {\it V} & 2003-2012 & 517 & 111.8 & 0.029 & 110.2 & 0.023 & 103.1 & 0.009 &  68.3 & 0.009 & 97.2 & 0.009 & \nodata & \nodata \\
18095 & D & {\it B} & 1993-2012 & 544 & 113.5 & 0.024 & 103.0 & 0.016 & 154.5 & 0.016 & 97.9 & 0.015 & 116.3 & 0.016 & \nodata & \nodata \\
18095 & E & {\it R$_C$} & 1998-2012 & 455 & 102.9 & 0.009 &  112.6 & 0.014 & 109.1 & 0.010 &  97.7 & 0.009 & 149.3 & 0.07 & \nodata & \nodata \\
18095 & F & {\it R$_C$} & 2003-2012 & 359 & 101.1 & 0.023 & 99.9 & 0.021 & \nodata & \nodata & \nodata & \nodata & \nodata & \nodata & \nodata & \nodata \\
\\
19386 & A & {\it V} & 1993-2008 & 412 & 101.8 & 0.026 & 98.6 & 0.025 & 103.5 & 0.024 & 106.3 & 0.018 & 89.2 & 0.010 & \nodata & \nodata \\
19386 & B & {\it V} & 1993-2000 & 170 &  99.5 & 0.035 & 91.7 & 0.018 & 109.3 & 0.012 & \nodata & \nodata & \nodata & \nodata & \nodata & \nodata \\
19386 & C & {\it V} & 2001-2008 & 242 & 103.2 & 0.045 & 98.8 & 0.031 & 106.7 & 0.021 & 88.3 & 0.012 & \nodata & \nodata & \nodata & \nodata \\
19386 & D & {\it R$_C$} & 2002-2007 & 100 & 100.4 & 0.041 & 105.1 & 0.022 & \nodata & \nodata& \nodata & \nodata & \nodata & \nodata & \nodata & \nodata \\
\\
19475 & A & {\it V} & 1994-2012 & 611 & 40.95 & 0.016 & 38.85 & 0.013 & 48.09 & 0.011 & \nodata & \nodata & \nodata &  \nodata &  \nodata &  \nodata \\
19475 & B & {\it R$_C$} & 1998-2012 & 403 & 40.98 & 0.016 & 38.84 & 0.013 & 48.18 & 0.011 &  \nodata & \nodata & \nodata &  \nodata &  \nodata &  \nodata \\
19475 & C & {\it V} & 2002-2012 & 456 & 40.99 & 0.018 & 38.79 & 0.016 &  48.11 & 0.014 &  \nodata & \nodata & \nodata &  \nodata &  \nodata &  \nodata \\
19475 & D & {\it R$_C$} & 2002-2012 & 339 & 41.02 & 0.017 & 38.80 & 0.013 & \nodata &  \nodata &  \nodata & \nodata & \nodata &  \nodata &  \nodata &  \nodata \\
19475 & E & {\it V} & 1994-1999 & 155 & 36.72 & 0.016 & 36.13 & 0.015 &   \nodata & \nodata &  \nodata & \nodata & \nodata &  \nodata &  \nodata &  \nodata \\
\enddata
\tablenotetext{a}{The formal, least-squares uncertainties determined using Period04 for each object in period {\it P} and amplitude {\it A} are approximately as follows:\\
IRAS 17436: $\pm$0.01 to $\pm$0.025 days for subsets A to D and $\pm$0.05 to $\pm$0.20 days for the shorter subsets E to G, with $\pm$0.001 to $\pm$0.002 mag, respectively; \\
IRAS 18095: $\pm$0.1 to $\pm$0.3 days for subsets A to E and $\pm$0.8 days for subset F (strong beat periods), with $\pm$0.001 to $\pm$0.002 mag, except for subset F, in which they are much more uncertain; \\
IRAS 19386: $\pm$0.1 (subset A) to $\pm$0.35 (subsets B to D) days, with $\pm$0.002 mag; \\
IRAS 19475: $\pm$0.01 to $\pm$0.04 days for subsets A to D and $\pm$0.07 days for the shorter subset E, with $\pm$0.002 mag, respectively. }
\tablenotetext{b}{A similarly good fit is achieved with P$_1$=98.5 days, A$_1$=0.013 mag and P$_2$=102.8 days, A$_2$= 0.012 mag}
\end{deluxetable}

\clearpage

\tablenum{8}
\begin{deluxetable}{rrrrlclccll}
\rotate \tablenum{8} 
\tablecaption{Results of the Period and Light Curve Study of Four Oxygen-Rich PPNs\label{results}}
\tabletypesize{\footnotesize}
\tablewidth{0pt} \tablehead{ \colhead{IRAS ID} &\colhead{P$_1$}
& \colhead{P$_2$/P$_1$} &\colhead{P$_{\rm prev}$\tablenotemark{a}} 
&\colhead{SpT} &\colhead{T$_{\rm eff}$\tablenotemark{b}} 
&\colhead{$\Delta$V\tablenotemark{c}}&\colhead{$\Delta$(V$-$R$_C$)\tablenotemark{b}}
&\colhead{$\Delta$(B$-$V)\tablenotemark{b}}&\colhead{$\Delta$T$_{\rm eff}$\tablenotemark{b}}&\colhead{Comments}\\
 &\colhead{(day)} 
 & &\colhead{(day)} & &\colhead{(K)} &\colhead{(mag)} &\colhead{(mag)} &\colhead{(mag)} &\colhead{(K)} &  } \startdata
17436+5003  & 45.2   & 1.06 & 40$-$46 & F3 Ib & 6600, 7100 & 0.21 & 0.065 & 0.095 & 700 & \nodata  \\
18095+2704  & 113.3  & 0.86 & 109 & F3 Ib & 6500 & 0.14 & 0.060 & 0.080 & 590 &  brightening trend \\
19386+0155  & 101.8  & 0.97 & 102, 98 & F5 Ib & 6800 & 0.20 & 0.070 & 0.105 & 770 &  brightening trend \\
19475+3119  & 41.0   & 0.95 & 43, 45 & F3 Ib & 7750, 7200 & 0.19 & 0.045 & 0.055& 400 & \nodata  \\
\enddata
\tablenotetext{a}{Previously published periods: IRAS 17436+5003: Fernie, series of papers; IRAS 18095+2704: \citet{ark10}; IRAS 19386+0155: \citet{ark10}; IRAS 19475+3119: \citet{ark06}.}
\tablenotetext{b}{T$_{\rm eff}$ determined from model atmosphere analyses as cited in Table~\ref{object_list}}
\tablenotetext{c}{The maximum brightness and color range observed in a season and the corresponding maximum temperature change based on $\Delta$(B$-$V) and color$-$temperature calibration of \citet[Table 15.7]{cox00}. }
\end{deluxetable}

\clearpage

\end{document}